\documentclass[a4paper,11pt]{article}
\pdfoutput=1
\usepackage{jheppub,MnSymbol}

\newcommand{\beq}{\begin{equation}}
\newcommand{\eeq}{\end{equation}}
\newcommand{\bea}{\begin{eqnarray}}
\newcommand{\eea}{\end{eqnarray}}

\newcommand{\ls}{\lambda_{S}}
\newcommand{\ld}{\lambda_{D}}
\newcommand{\lp}{\lambda_{D}^{\prime}}
\newcommand{\lpp}{\lambda_{D}^{\prime \prime}}
\newcommand{\mss}{M_{{S,S}}}
\newcommand{\msd}{M_{{D,S}}}
\newcommand{\ps}{\phi_{S}}
\newcommand{\pd}{\phi_{D}}
\newcommand{\pmi}{\phi^+}
\newcommand{\pr}{\phi_{R}}
\newcommand{\pim}{\phi_{I}}
\newcommand{\de}{\psi_{D_1}}
\newcommand{\dz}{\psi_{D_2}}
\newcommand{\mfs}{M_{S,F}}
\newcommand{\mfd}{M_{D,F}}
\newcommand{\ye}{{y}_1}
\newcommand{\yz}{{y}_2}
\newcommand{\s}{\psi_S}

\begin{document}

\preprint{MS-TP-18-05}

\title{A singlet doublet dark matter model with radiative neutrino masses}

\author[a]{Sonja Esch}
\author[a]{\!\!, Michael Klasen}
\author[b]{\!\!, Carlos E. Yaguna}

\affiliation[a]{Institut f\"ur Theoretische Physik,
 Westf\"alische Wilhelms-Universit\"at
 M\"unster, Wilhelm-Klemm-Stra\ss{}e 9, D-48149 M\"unster, Germany}
\affiliation[b]{Universidad Pedag\'ogica y Tecnol\'ogica de Colombia
 at Sede Central Tunja-Bayac\'a-Colombia, Avenida Central del Norte
 39-115}

\emailAdd{sonja.esch@uni-muenster.de}
\emailAdd{michael.klasen@uni-muenster.de}
\emailAdd{cyaguna@gmail.com}

\date{\today}

\abstract{We present a detailed study of a combined singlet-doublet
 scalar and singlet-doublet fermion model for dark matter. These
 models have only been
 studied separately in the past. We show that their combination
 allows for the radiative generation of neutrino masses, but that
 it also implies the existence of lepton-flavour violating (LFV)
 processes. We first analyse the dark matter, neutrino mass and LFV
 aspects separately. We then perform two random scans for scalar
 dark matter imposing Higgs mass, relic density and neutrino mass
 constraints, one over the full parameter space, the other over
 regions where scalar-fermion
 coannihilations become important. In the first case, a large part
 of the new parameter space is excluded by LFV, and the remaining
 models will be probed by XENONnT. In the second case, direct
 detection cross sections are generally too small, but a substantial
 part of the viable models will be tested by future LFV experiments.
 Possible constraints from the LHC are also discussed.
}


\maketitle
\flushbottom

\section{Introduction}
\label{sec1}

While the Standard Model (SM) is certainly a highly successful theory of
particle physics and, with the discovery of a Higgs-like boson by the
ATLAS \cite{Aad:2012tfa} and CMS \cite{Chatrchyan:2012xdj} experiments
at the CERN Large Hadron Collider (LHC) in 2012, is sometimes presented
as complete, it is generally believed to suffer from conceptional and
phenomenological deficiencies. Important examples are the unexplained
large hierarchy of interactions \cite{Georgi:1974yf,Veltman:1980mj}, the
surprisingly small, but non-zero masses of neutrinos
\cite{Gonzalez-Garcia:2014bfa} established first in 1998 by the
oscillations of atmospheric \cite{Fukuda:1998mi}, then of solar
\cite{Ahmad:2001an,Ahmad:2002jz} and reactor neutrinos
\cite{Araki:2004mb}, and the observational evidence from many different
scales \cite{Klasen:2015uma} for a sizeable dark matter density in the
Universe, quantified by WMAP \cite{Hinshaw:2012aka} and then, more
precisely, by the Planck satellite \cite{Ade:2013zuv} to be
$\Omega^{\rm obs} h^2=0.1186\pm0.0031$. Here, $h$ denotes the present
Hubble expansion rate in units of 100 km s$^{-1}$ Mpc$^{-1}$.

An intriguing remedy for these problems is to extend the SM minimally
by a small number ($\leq 4$) of new scalar and fermion fields, which
allow to generate small neutrino masses radiatively and which have one
(or more) neutral components that could represent dark matter. One of
the most popular so-called radiative seesaw models is the scotogenic
model with only one additional scalar (inert Higgs) SU(2)$_L$ doublet
and a (right-handed neutrino) fermion singlet \cite{Ma:2006km}, for
which we recently demonstrated the importance of coannihilations
between the scalar dark matter and the right-handed neutrinos
\cite{Klasen:2013jpa}. Many variants with general $N$-tuplets
containing in general two scalars and one fermion have subsequently
been proposed \cite{Ma:2008cu,Farzan:2010mr,Aoki:2011yk,Law:2013saa}.
The new fields are usually assumed to be odd under a Z$_2$ symmetry
to stabilize the dark matter and (in some cases) to prevent tree-level
contributions to neutrino masses. All one-loop models connecting
neutrino masses to dark matter with at most four additional fields
have been classified \cite{Restrepo:2013aga} following the notation of
a systematic study of the $d=5$ Weinberg operator at one-loop order
\cite{Bonnet:2012kz}. The classification has recently been extended to
two loops \cite{Simoes:2017kqb} following Ref.\ \cite{Sierra:2014rxa}
(cf.\ also Sec.\ 4.7 in \cite{Cai:2017jrq}).

Apart from the models mentioned above, which all belong to the one-loop
topology T3 as defined in Refs.\ \cite{Bonnet:2012kz,Restrepo:2013aga},
several models with other topologies have also been studied in the past.
They include a model of topology T1-1 with three scalars, two of which
are equivalent, and one fermion \cite{Farzan:2009ji}, and a model of
topology T1-3 with one scalar and three fermions
\cite{Fraser:2014yha,Restrepo:2015ura,Klasen:2016vgl}. In both cases,
the lightest scalar was assumed to represent the dark matter,
constraints from the Higgs and neutrino sectors and the relic density
were imposed, and their phenomenology, in particular of lepton-flavour
violating processes, has been studied. In this paper, we study a model
of topology T1-2 with two scalars and two fermions. A general
discussion of this topology, without establishing the particle content
of specific models, has been presented in Ref.\ \cite{Farzan:2012ev}.
To complete the list, note that models of topology T2 can be discarded
on dimensional arguments, while models of T4, T5 and T6 always have
tree-level contributions to neutrino masses and thus no obvious
connection to dark matter. In addition, three out of six models of 
type T4 and all models of type T5 and T6 are divergent at one loop.

As mentioned above, we study in this paper a model of topology T1-2
with two scalars and two fermions. We focus on the first of these
models, T1-2-A in the classification of Ref.\ \cite{Restrepo:2013aga},
with a singlet Majorana fermion $\psi_S$ of hypercharge parameter
$\alpha=0$, that determines also the hypercharges of the vector-like
fermion doublet $\psi_D$, the singlet scalar $\phi_S$, and the complex
doublet scalar $\phi_D$. The alternative case $\alpha=-2$, called the
Inert Zee Model, has previously been analysed and found to be
consistent with constraints from neutrino oscillation, dark matter and
lepton-flavour violation data \cite{Longas:2015sxk}. A doublet-triplet
model, T1-2-F with $\alpha=-1$, has also been analysed in a similar way
\cite{Betancur:2017dhy}. Since our model T1-2-A with $\alpha=0$ is
equivalent to T1-2-C with $\alpha=-1$, our results equally apply to
this model. It is furthermore one of the models, together with the
model T1-3-A and $\alpha=0$ \cite{Fraser:2014yha,Restrepo:2015ura,%
Klasen:2016vgl}, that allows for gauge coupling unification at a
scale $\Lambda={\cal O}(10^{13})$ GeV, at variance with most of the
other models mentioned above including the original scotogenic model
\cite{Hagedorn:2016dze}.

The paper is organised as follows: In Sec.\ \ref{sec:2} we define
our model by describing its particle content, mass and interaction
Lagrangians, free parameters and particle mixings. In Sec.\
\ref{sec:3} we review the implications for the dark matter relic
density first of the scalar and fermion sectors separately, before
we study the new regions of parameter space that open up when they
couple to each other as required for neutrino mass generation.
The latter is discussed in Sec.\ \ref{sec:4} in analytic and
numerical form using a Casas-Ibarra parameterisation and limits
of small couplings. As lepton-flavour violation always occurs in
this type of models, we compute the branching ratios for $\mu\to
e\gamma$ and similar processes in Sec.\ \ref{sec:5}. Sec.\
\ref{sec:6} contains our main numerical results from two random
scans, one over the full parameter space, one in the coannihilation
region of scalars and fermions. We impose all current experimental
constraints, estimate the senstivity of future experiments and
and discuss the possible impact of the LHC. Our conclusions are
given in Sec.\ \ref{sec:7}.

\section{Particle content and interactions in the model}
\label{sec:2}

In our model T1-2-A, the SM is extended by two scalars, a real singlet
$\phi_S$ and a complex doublet $\phi_D$, and two fermions, a
Weyl singlet $\psi_S$ and a vector-like doublet $\psi_D$, which can be
decomposed into two chiral Weyl fermions $\de$, $\dz$. The new (SM)
particles are all assumed to be odd (even) under a discrete Z$_2$
symmetry to render dark matter stable and generate neutrino masses only
at the loop level. Under the combined SU(2)$_L\otimes$U(1)$_Y\otimes$Z$_2$
symmetry group, the new particles carry the quantum numbers
\bea
 \ps \propto (1,0,-) &,&
 \pd = \left(\begin{array}{c} \pmi \\ \pr + i \pim \end{array}\right)
  \propto (2,\frac{1}{2},-) \\
 \s \propto (1,0,-) &,&
 \de = \left( \begin{array}{c} \psi^0_{D_{1,L}} \\ \psi^-_{D_{1,L}} \end{array}\right)
  \propto (2, -\frac{1}{2}, -) ~,~
 \dz = \left(\begin{array}{c} -(\psi^-_{D_{2,R}})^{\dagger} \\ (\psi^0_{D_{2,R}})^{\dagger} \end{array}\right)
 \propto (2, \frac{1}{2}, -).
\eea
The pair of Weyl fermions are required for the cancellation of
anomalies. For the doublets, the decompositions according to the
physical charges $Q=T_3+Y/2$ have also been indicated. It is evident
that the model can be seen as a combination of the singlet-doublet
scalar \cite{Cohen:2011ec,Cheung:2013dua,
Banik:2014cfa,
Cabral-Rosetti:2017mai} 
and the singlet-doublet fermion dark matter model
\cite{Cohen:2011ec,Cheung:2013dua,
Calibbi:2015nha,
Banerjee:2016hsk,
Abe:2017glm}, 
which have both well been studied separately. Our model allows in
addition for the radiative generation of neutrino masses.

The Lagrangian of the singlet-doublet scalar model
\bea
 -\mathcal{L}_{\text{scalar}} &=&
 \frac{1}{2} \ls \ps^2 |H|^2 + \ld |\pd|^2 |H|^2 + \lp |\pd^{\dagger}H|^2
 + \frac{1}{2} \lpp \left[\left(\pd^{\dagger}H\right)^2+\text{h.c.}\right]
 \nonumber \\
 &+& A \left[\pd^{\dagger} H \ps + \text{h.c.}\right]
 + \frac{1}{2} \mss^2 \phi_S^2 + \msd^2 |\phi_D|^2
 \label{scallag}
\eea
involves in general five different couplings of the new scalars with the
complex SM Higgs doublet $H$, which acquires a vacuum expectation value
$v=246$ GeV and thus breaks the electroweak symmetry. One of these
couplings ($A$) links the singlet and doublet scalars with the SM
Higgs in a Yukawa-like form. The Z$_2$ symmetry remains unbroken,
{\it i.e.} the new scalars do not have quartic potentials and do not
acquire a vacuum expectation value, but have explicit mass terms.
The Lagrangian of the singlet-doublet fermion model
\beq
 -\mathcal{L}_{\text{fermion}} = \ye \de H \s + \yz \dz H^\dagger \s
 + \frac{1}{2} \mfs \s^2 + \mfd \de \dz + \text{h.c.}
 \label{fermlag}
\eeq
contains a Majorana mass term for the singlet $\s$, a Dirac mass term
for the doublet $\psi_D$ as well as two Yukawa terms connecting the
singlet and doublet fermions with the SM Higgs.
The presence of scalar and fermion singlets and doublets allows also to
define couplings of the new particles to the SM leptons,
\beq
 -\mathcal{L}_{\text{lepton}} = g_{1i}  L_i \ps \dz  + g_{2i}  L_i\pd \s + \text{h.c.},
 \label{neutlag}
\eeq
where
\begin{align}
 L_i = \left( \begin{array}{c} \nu_{i,L}^0\\ e_{i,L}^-\end{array}\right) \propto (2,-\frac{1}{2},+)
\end{align}
is the SM lepton doublet of generation $i$. These new Yukawa couplings
$g_{ij}$ allow to obtain one-loop neutrino masses after electroweak
symmetry breaking, while at tree level the neutrinos remain massless.
A term $g_{3i}e_{i,R}^c\phi_D^\dagger\psi_{D_1}$ with no effect on
neutrino masses has been omitted.
The presence of these non-zero neutrino masses implies lepton flavour
violation. Inspection of Eqs.\ (\ref{scallag})--(\ref{neutlag}) shows
that the terms featuring $A$, $\ye$, $\yz$ and $g_{ij}$ do not allow
to determine the lepton numbers of the new particles consistently
without violating lepton flavour. Only when either $g_{ij}$ or all
mixing terms $A$, $\ye$ and $\yz$ vanish, lepton flavour symmetry can
be restored.

\begin{table}[t]
 \caption{Free parameters in model T1-2-A.}
 \label{paramTab}
 \begin{center}
  \begin{tabular}{|l|l|}
    \hline
    Sector			& Parameters \\
    \hline
    Scalar sector		&  $\mss$, $\msd$, $\ls$, $\ld$, $\lp$, $\lpp$, $A$ \\
    Fermion sector	&  $\mfs$, $\mfd$, $\ye$, $\yz$ \\
    Neutrino sector 	& $g_{11}$, $g_{12}$, $g_{13}$, $g_{21}$, $g_{22}$, $g_{23}$ \\
    \hline
  \end{tabular}
 \end{center}
\end{table}
In the basis of real neutral scalars ($\phi_S$, $\phi_R$, $\phi_I$), one
can extract their mass matrix after electroweak symmetry breaking
from the Lagrangian in Eq.\ (\ref{scallag}) \cite{Cheung:2013dua}
\bea
 M^2_{\text{scalar}} &=& 
 \left( 
 \begin{array}{c c c} 
 \mss^2 + \frac{1}{2} v^2 \ls & A v                                         & 0 \\
 A v                          & \msd^2 + \frac{1}{2} v^2(\ld + \lp + \lpp)  & 0 \\
 0 			      & 0 					    & \msd^2 + \frac{1}{2} v^2(\ld+\lp-\lpp)
 \end{array} 
 \right).\quad
\eea
In the absence of new sources of CP-violation, $\phi_I$ does not mix
with either $\phi_R$ or $\phi_S$.
The mass matrix for the neutral fermions can similarly be derived from
the Lagrangian in Eq.\ (\ref{fermlag}) \cite{Cheung:2013dua}, so that
\bea
 M_{\text{fermion}} &=&
 \left(
 \begin{array}{ccc}
  \mfs          & m_y\cos\theta & m_y\sin\theta \\
  m_y\cos\theta & 0             & \mfd \\
  m_y\sin\theta & \mfd          & 0 
 \end{array}
 \right).
\eea
Here, $m_y = \frac{1}{\sqrt{2}} y v$ with $y=\sqrt{y_1^2+y_2^2}$,
$v$ is the SM Higgs vacuum expectation value, and $\tan\theta=\frac{\yz}
{\ye}$, so that $\ye=y\cos\theta$ and $\yz=y\sin\theta$. Mixing of
the neutral scalar and fermion particles is parameterized by the
matrices $U_S$ and $U_F$, which leads to the mass eigenstates
\bea
 \left( \begin{array}{c} X_1 \\ X_2 \\ X_3 \end{array} \right) =
 U_S  \left( \begin{array}{c} \ps \\ \pr \\ \pim \end{array} \right)  &\quad,\quad&
 \left( \begin{array}{c} \chi_1 \\ \chi_2 \\ \chi_3 \end{array} \right) =
 U_F  \left( \begin{array}{c} \s  \\ \de^0 \\ \dz^0  \end{array} \right).
 \label{neutralStates}
\eea
Due to the mixing with $\psi_S$, all physical fermions $\chi_i$ are
Majorana particles.
In addition to the neutral sector, one obtains a charged scalar $\pd^-$
with mass $m_{\pd^-}$ and a charged fermion
\beq
 \chi^- = \left( \begin{array}{c} D_{1,L}^-  \\ D_{2,R}^- \end{array} \right)
\eeq
with mass $m_{\psi^-} = \mfd$.
In total, our model contains 17 free parameters, which are summarized
in Tab.\ \ref{paramTab}.

\section{Dark matter relic density}
\label{sec:3}

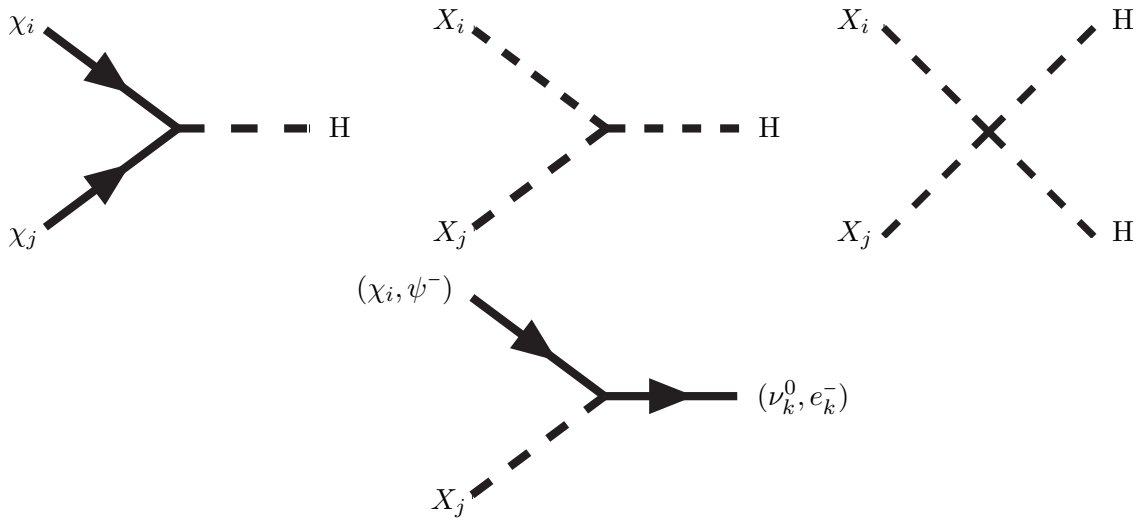
\begin{figure}
\begin{minipage}{0.20\textwidth}\begin{picture}(310,100)(0,0)
    \put(5,5){\mbox{\resizebox{!}{3cm}{\includegraphics{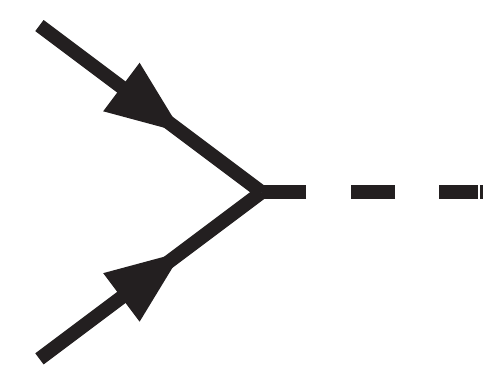}}}}
    \put(0,85){$ \chi_i$}
    \put(0, 5){$\chi_j$}
    \put(120,44){H}
\end{picture}\end{minipage}\hspace*{2.5cm}
\begin{minipage}{0.20\textwidth}\begin{picture}(310,100)(0,0)
    \put(5,5){\mbox{\resizebox{!}{3cm}{\includegraphics{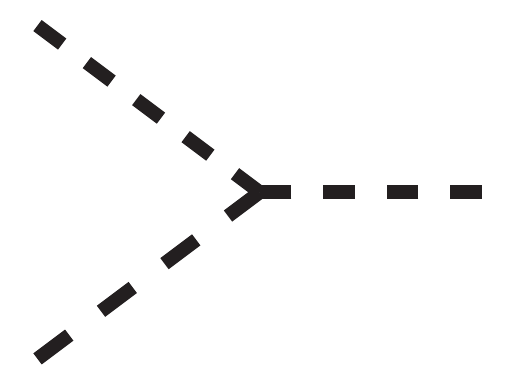}}}}
    \put(-2,85){$X_i$}
    \put(-2, 5){$X_j$}
    \put(120,44){H}
\end{picture}\end{minipage}\hspace*{-3cm}
\begin{minipage}{0.20\textwidth}\begin{picture}(310,100)(0,0)
    \put(155,5){\mbox{\resizebox{!}{3cm}{\includegraphics{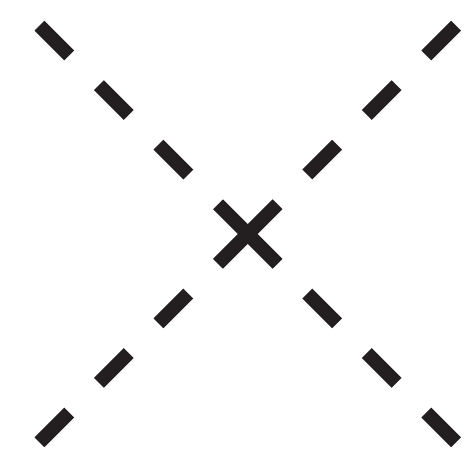}}}}
    \put(145,85){$ X_i$}
    \put(145, 5){$X_j$}
    \put(248,85){H}
    \put(248, 5){H}    
\end{picture}\end{minipage}
\\ \hspace*{5.5cm}
\begin{minipage}{0.20\textwidth}\begin{picture}(310,100)(0,0)
    \put(5,5){\mbox{\resizebox{!}{3cm}{\includegraphics{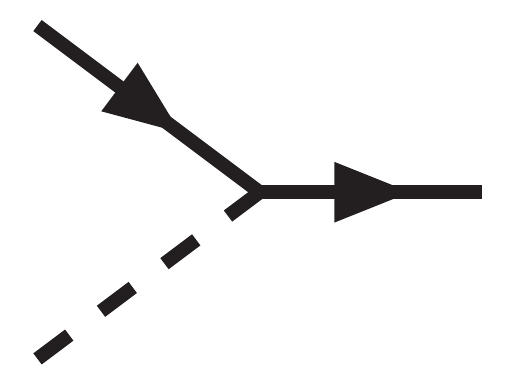}}}}
    \put(-30,85){($\chi_i,\psi^-$)}
    \put(-2, 5){$X_j$}
    \put(120,44){($\nu_k^0,e_k^-$)}
\end{picture}\end{minipage}
\caption{New vertices that link the Z$_2$-odd sector to the Standard Model.}
\label{fig:1}
\end{figure}

Let us first review the fermion and scalar dark matter sectors of our
model separately. The singlet-doublet fermion model has been shown to be
severely constrained by LUX and XENON1T. Only regions with small Yukawa
coupling $y<0.1$ are still allowed, and those are fine-tuned to at least
10\%. The singlet-doublet scalar model was less constrained by LUX. In
the limit $\lambda=\lambda_S=\lambda_D$ with $\lambda_D'=\lambda_D''=0$,
regions of $\lambda<0$ were, however, excluded by XENON1T, while
$\lambda>0$ remained viable, albeit again with a typical fine-tuning of
$\leq10$\% \cite{Cheung:2013dua}. The correct relic density is reached
in a thermal freeze-out through electroweak interactions and often
through resonances or via coannihilation processes \cite{Cohen:2011ec}.
In both cases, the dark matter mass had to be in the few hundred GeV to
few TeV range. Relic density and direct detection constraints are
correlated through the same Higgs couplings, shown in the top row of
Fig.\ \ref{fig:1}.
We have verified that we correctly reproduce the results in
Ref.\ \cite{Cheung:2013dua} separately for fermion and scalar
singlet-doublet dark matter.

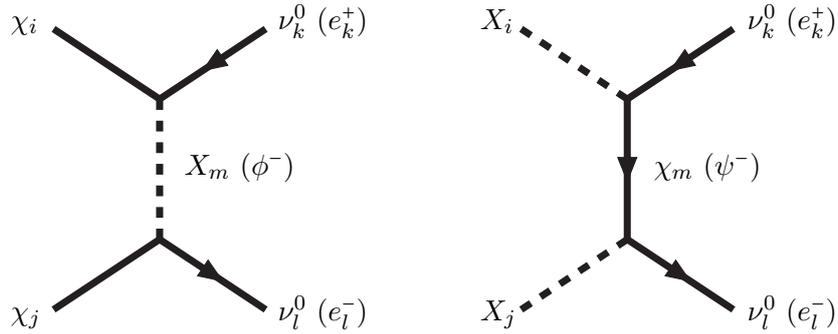
\begin{figure}
\begin{center}
\begin{minipage}{0.4\textwidth}\begin{picture}(210,150)(0,0)
    \put(30,5){\mbox{\resizebox{!}{4cm}{\includegraphics{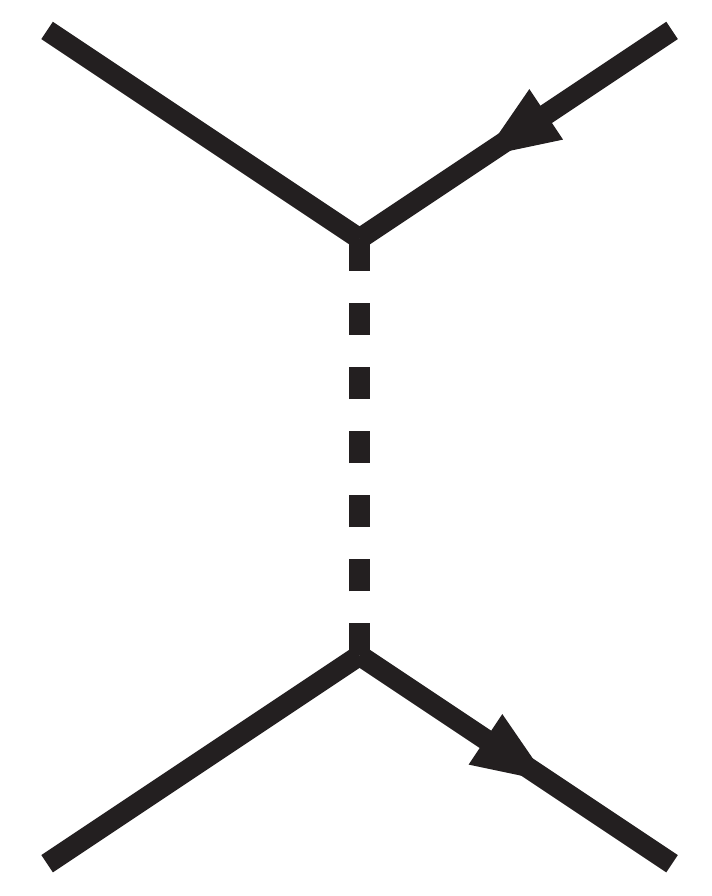}}}}
    \put(20,115){$\chi_i$}
    \put(20, 5){$\chi_j$}
    \put(85, 60){$X_m$ ($\phi^-$)} 
    \put(120,115){$\nu_{k}^0$ ($e^+_k$)}
    \put(120, 5){$\nu_{l}^0$ ($e^-_l$)}    
\end{picture}\end{minipage}
\begin{minipage}{0.4\textwidth}\begin{picture}(210,150)(0,0)
    \put(30,5){\mbox{\resizebox{!}{4cm}{\includegraphics{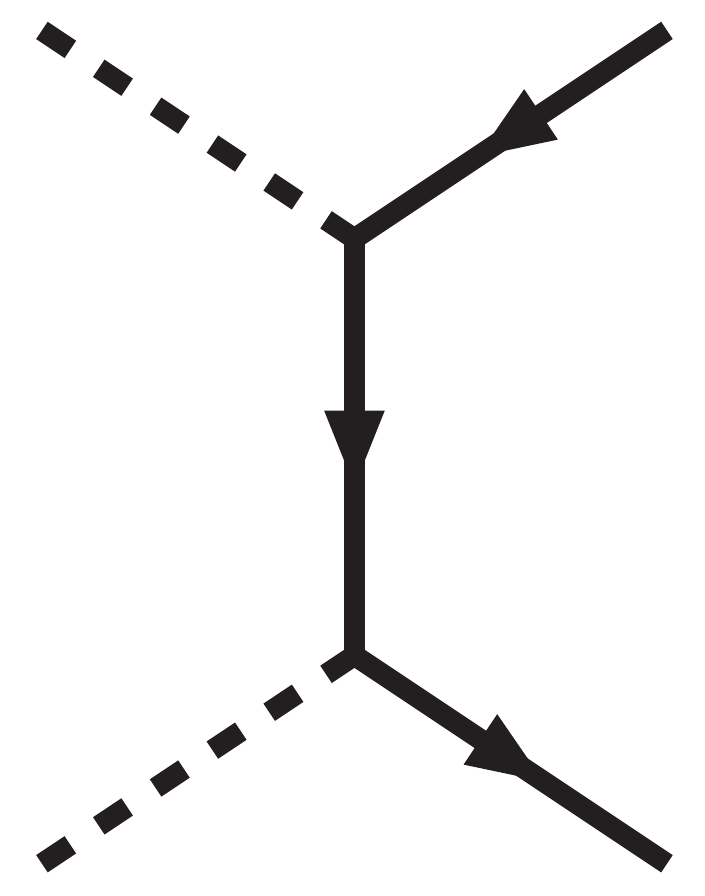}}}}
    \put(20,115){$X_i$}
    \put(20, 5){$X_j$}
    \put(85, 60){$\chi_m$ ($\psi^-$)} 
    \put(120,115){$\nu_{k}^0$ ($e^+_k$)}
    \put(120, 5){$\nu_{l}^0$ ($e^-_l$)}    
\end{picture}\end{minipage}
\end{center}
\caption{(Co-)annihilation processes of fermion (left) and scalar
 (right) dark matter particles to SM leptons.}
\label{fig:2}
\end{figure}

Our model allows for both fermion and scalar dark matter candidates
as well as for fermion-scalar coannihilation processes,
mediated by the lepton vertex shown in the bottom row of Fig.\
\ref{fig:1}. This implies also that our dark sector can be leptophilic,
if the Higgs couplings are small, and
that new annihilation processes, shown in Fig.\ \ref{fig:2}, can
occur, which will in general
relax the correlation of relic density and direct detection constraints
via the Higgs couplings. The annihilation cross sections of these
processes scale with the relative velocity $v$ of the thermal relic as
\bea
 \sigma v (\chi_i \chi_j \rightarrow e^+_k e^-_l) \propto v^2 &\ ,\ &
 \sigma v (X_i X_j \rightarrow e_k^+ e^-_l) \propto v^4,\\
 \sigma v (\chi_i \chi_j \rightarrow \nu^0_k \nu^0_l) \propto v^0&\ ,\ &
 \sigma v (X_i X_j \rightarrow \nu^0_k \nu^0_l) \propto v^2,
\eea
respectively. Conversion processes $\chi_i\chi_j
\leftrightarrow X_k X_l$ can also occur, so that the evolution of the
dark matter
particle is described by a set of coupled Boltzmann equations. We solve
these with micrOMEGAs 4.0.3 \cite{Belanger:2014vza} after implementation
of the Lagrangians in Eqs.\ (\ref{scallag})--(\ref{neutlag}) in LanHEP
\cite{Semenov:2014rea} and SARAH 4 \cite{Staub:2013tta} as a cross-check
and to facilitate use of the spectrum generator SPheno 3.3.6
\cite{Porod:2011nf}.

\begin{figure}
\begin{center}
\includegraphics[width=0.95\textwidth]{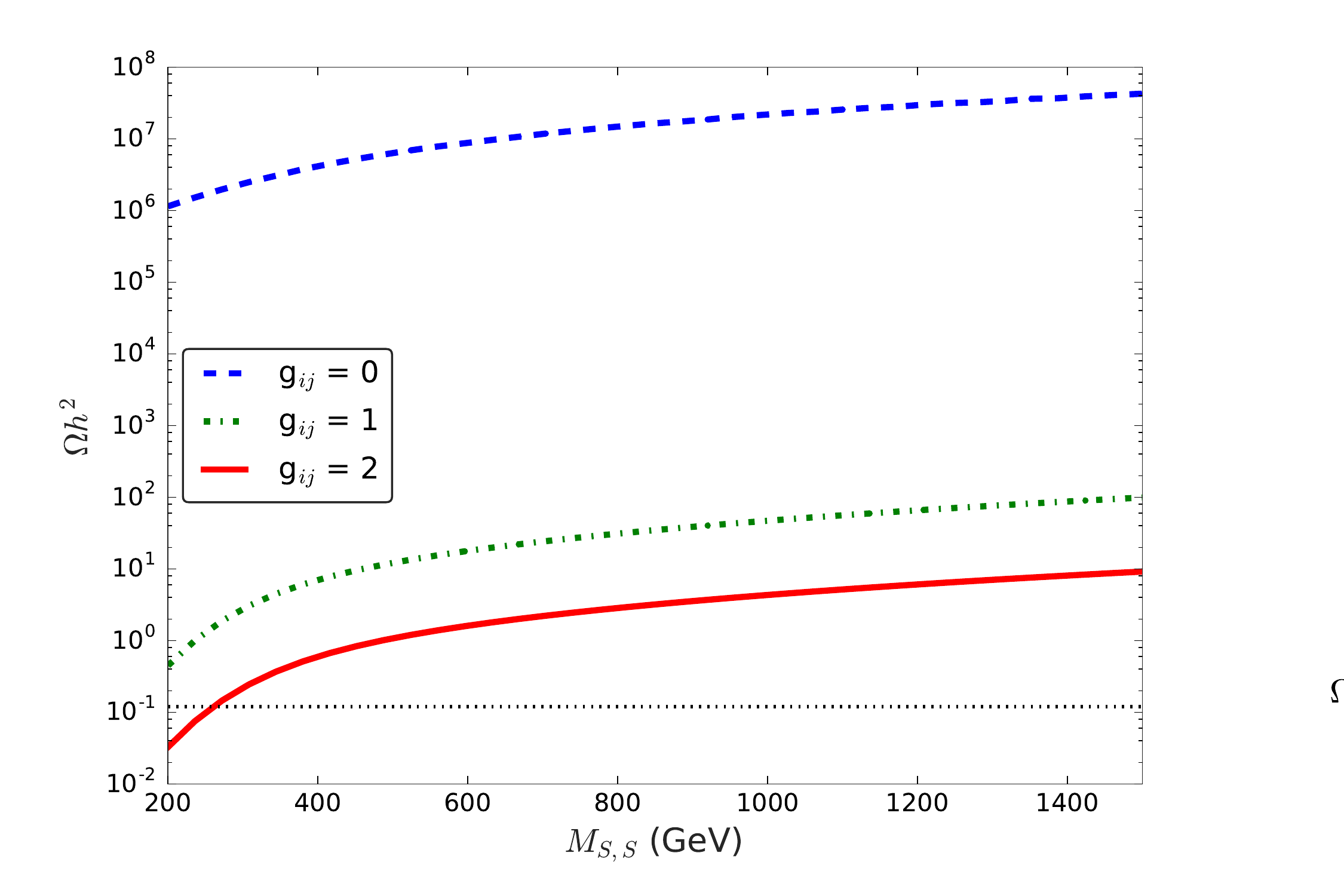}
\end{center}
\caption{Influence of the scalar-fermion couplings $g_{ij}$ on the
 singlet scalar dark matter relic density as a function of its mass
 parameter $\mss$.}
\label{fig:3}
\end{figure}

In Fig.\ \ref{fig:3}, we demonstrate the impact of the couplings 
$g_{ij}$ among the new fermion and scalar sectors on the singlet scalar
dark matter relic density as function of its mass. The other masses 
$M_{D,S}$, $M_{S,F}$ and $M_{D,F}$ are scaled to it and larger by
factors of 3, 3.1 and 2.5, respectively, so that coannihilations are
unimportant. Mixings and (co-)annihilations in the scalar sector alone
are further suppressed by small scalar couplings $\lambda_S$ etc.,
which are all set to values below 10$^{-5}$, so that the relic
density with $g_{ij}=0$ (dashed blue line) is seven orders of magnitude
larger than the value observed by Planck (dotted black line).
Increasing the scalar-fermion coupling to $g_{ij}=2$ (full red line)
then allows to bring the singlet scalar relic density in agreement with
the observation, whereas $g_{ij}=1$ (dot-dashed green line) is not quite sufficient.
The direct detection cross section is independent of
$g_{ij}$, so that it can be decoupled from the relic density constraint.

\begin{figure}
\begin{center}
\includegraphics[width=0.95\textwidth]{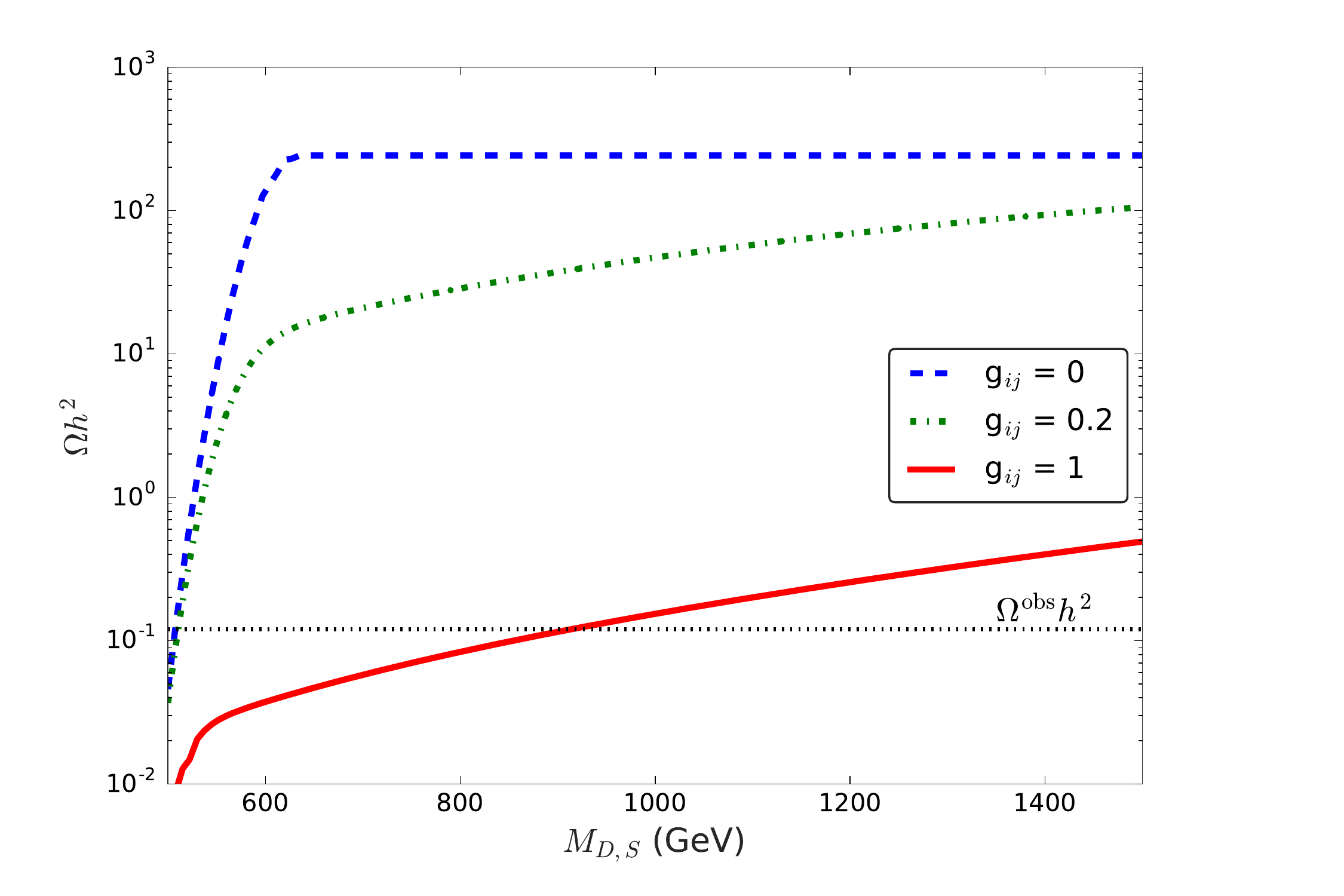}
\end{center}
\caption{Influence of the scalar-fermion couplings $g_{ij}$ on the
 singlet fermion dark matter relic density as function of the doublet
 scalar mass parameter $\msd$.}
\label{fig:4}
\end{figure}

Next, we consider the effect of the couplings $g_{ij}$ on singlet
fermion dark matter. Fig.\ \ref{fig:4} shows its relic density as a
function of the doublet scalar mass $\msd$. The singlet and doublet
fermion masses have been fixed at $\mfs=450$ GeV and $\mfd=3$ TeV,
respectively, and the singlet scalar mass is 2.5 TeV, so
that dark matter is always fermionic in the mass range shown. Since the
Yukawa couplings $y_1=0.06$ and $y_2=0.28$ are small, it is also
dominated by the singlet component. In the absence of scalar-fermion
couplings $g_{ij}$ (dashed blue line), the relic density stays constant
above $M_{D,S}=600$ GeV. Below this value, the mass difference of the
lightest physical scalar and fermion falls to a few tens of GeV and
conversion processes $\chi_i\chi_j\leftrightarrow X_kX_l$
can occur, which depletes the relic density to acceptable values when
$\msd\simeq\mfs$. When the scalar-fermion couplings are set to $g_{ij}=0.2$
(dot-dashed green line) or even to $g_{ij}=1$ (full red line), the
relic density falls monotonically with the doublet scalar mass, as coannihilations
become more and more important. Even for large values of $\msd$, the
relic density is smaller than for $g_{ij}=0$ due to the presence of
additional annihilation channels into SM neutrinos and charged leptons
(see Fig.\ \ref{fig:2} left). Their effect becomes smaller due to the
scalar propagator suppression as $\msd$ increases. The direct detection
cross section is, of course, again independent of $g_{ij}$.

\begin{figure}
\begin{center}
\includegraphics[width=0.93\textwidth]{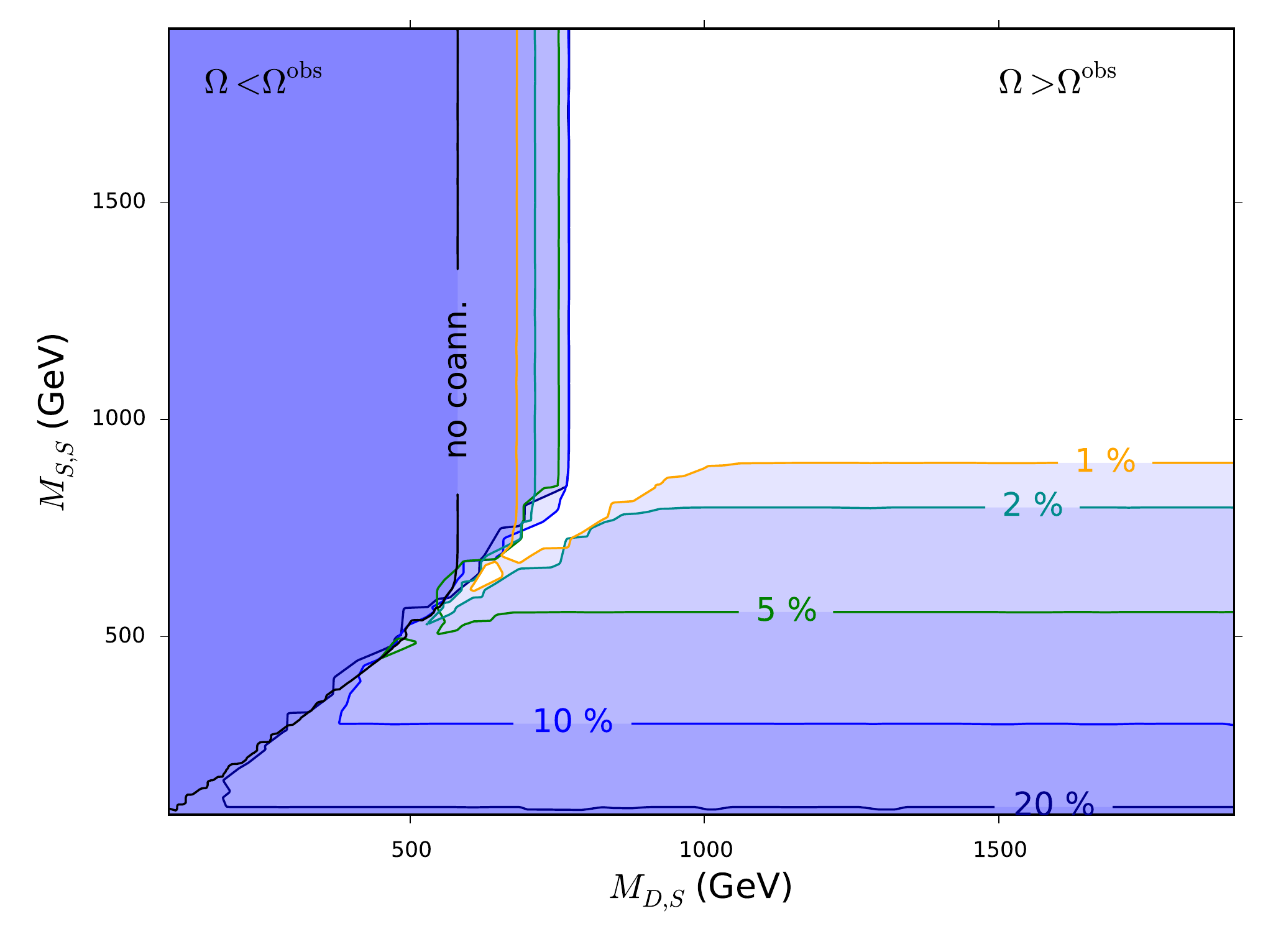}
\end{center}
\caption{Viable and excluded values of the scalar dark matter relic
 density in the mass plane $\msd$--$\mss$ in the presence of similarly
 light singlet or doublet fermions with couplings $g_{ij}=0.75$.}
 \label{fig:5}
\end{figure}

In Fig.\ \ref{fig:5}, we analyse the scalar dark matter relic density
in the mass plane $\msd$--$\mss$ in the presence of similarly light
singlet or doublet fermions for a fixed, sizeable value of
$g_{ij}=0.75$. The scalar couplings are either small ($\ls=\lpp=
10^{-4}$, $A=10^{-4}$ GeV) or zero ($\ld=\lp=0$), so that the dark
matter particle is mostly a singlet when $\mss<\msd$ and a doublet when
$\msd<\mss$. The Yukawa couplings have been set to $y_1=0.3$ and $y_2=
0.2$, respectively. When the fermions are decoupled to $\mfs=\mfd=5$
TeV (full black line), coannihilations cannot take place, and one
recovers the result in Fig.\ 12 of Ref.\ \cite{Cheung:2013dua}.
However, when the relative mass difference of singlet fermions to
doublet scalars is in the few-percent range (coloured vertical lines),
the allowed range in the relic density is extended beyond $\msd=600$
GeV to about 750 GeV. Conversely, when the relative mass difference of
doublet fermions to singlet scalars is in the few-percent range
(coloured horizontal lines), a new region of singlet scalar dark matter
opens up due to scalar-fermion coannihilations, conversion processes
and annihilations into SM leptons. For relative mass differences of
1--10\%, the singlet scalar dark matter mass cannot exceed 900 and 250
GeV, respectively.

\section{Neutrino masses}
\label{sec:4}

The combination of a singlet-doublet scalar and a singlet-doublet
fermion sector in model T1-2-A allows for the radiative
generation of neutrino masses through the box diagram depicted in Fig.\
2 of Ref.\ \cite{Restrepo:2013aga} and thus for a natural explanation of
the small relative size of neutrino masses with respect to other fermion
masses. Radiative neutrino mass models have been reviewed extensively in
Ref.\ \cite{Cai:2017jrq} including both radiative Dirac and Majorana
mass schemes. The latter can be classified according to the loop-order
realization of the Weinberg operator. Here, we focus on the one-loop
realization of the $d=5$ operator \cite{Bonnet:2012kz}.

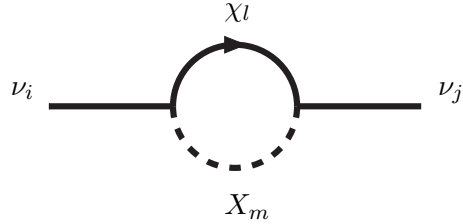
\begin{figure}
\begin{center}
\begin{minipage}{0.7\textwidth}\begin{picture}(300,100)(-30,-10)
    \put(50,0){\mbox{\resizebox{!}{0.18\textwidth}{\includegraphics{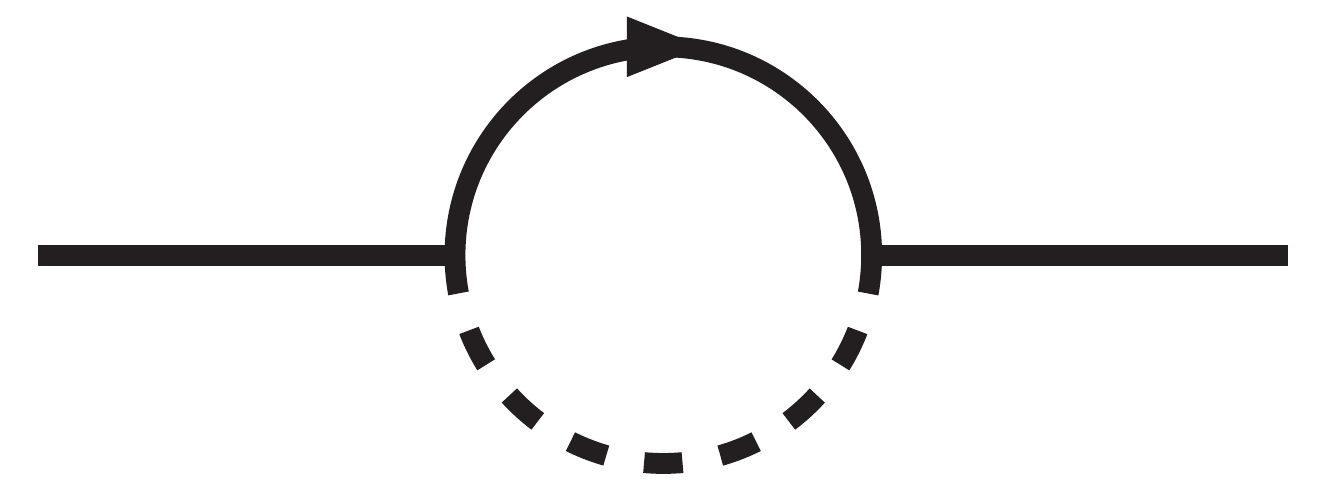}}}}
    \put( 40,30){$\nu_i$}
    \put(200,30){$\nu_j$}
    \put(120,60){$\chi_l$}
    \put(120,-15){$X_m$}
\end{picture}\end{minipage}
\end{center}
\caption{Neutrino-mass generation at one loop in model T1-2-A after
 electroweak symmetry breaking.}
\label{fig:6}
\end{figure}

After electroweak symmetry breaking, Majorana neutrino masses in our
model are generated at one loop by the diagram shown in Fig.\
\ref{fig:6}. The neutrino mass matrix is then given by
\bea
 M_{\nu,ij} &=& \sum_{l,m} \frac{1}{16\pi^2} \frac{M_{\chi_l}}{M_{X_m}^2-M_{\chi_l}^2}
 \left( M_{\chi_l^2} \ln M_{\chi_l}^2 - M_{X_m}^2 \ln M_{X_m}^2 \right)
\left[ U_{F, 3l}^2 U_{S, 1m}^2 g_{1i}g_{1j}\right.\nonumber \\
 && \left.
               + U_{F, 1l} U_{F, 3l} U_{S, 1m} U_{S, 2m} (g_{1i} g_{2j}+g_{1j} g_{2i})
               + U_{F, 1l}^2 (U_{S,2m}^2 - U_{S,3m}^2) g_{2i} g_{2j} \right],
 \label{neutmass}
\eea
which can be written as $M_{\nu} = g^T M g$ with the scalar-fermion coupling matrix
\beq
 g = \left( \begin{array}{ccc} g_{11} & g_{12 } & g_{13} \\ g_{21} & g_{22} & g_{23} \end{array} \right),
\eeq
the elements of the symmetric matrix $M$
\bea
  M_{11} & = &\sum_{l,m} m_{lm} U_{F,3l}^3 U_{S,1m}^2, \nonumber \\
  M_{12} & = &\sum_{l,m} m_{lm} U_{F,1l}U_{F,3l} U_{S,1m} U_{S,2m} ~=~ M_{21}, \\
  M_{22} & = &\sum_{l,m} m_{lm} U_{F,1l}^2 \left( U_{S,2m}^2 - U_{S,3m}^2 \right) \nonumber 
 \label{neutrino_coefficients}
\eea
and the mass function
\beq
  m_{lm} = \frac{1}{16\pi^2} \frac{M_{\chi_l}}{M_{X_m}^2-M_{\chi_l}^2} \left( M_{\chi_l^2} \ln M_{\chi_l}^2 - M_{X_m}^2 \ln M_{X_m}^2 \right).
 \label{mass_function}
\eeq
In the limit of vanishing $\lambda_S$, $\lambda_D$ etc. and small values of
$A,y_1,y_2\ll1$, the matrix elements of $M$ can be expanded and expressed as
\bea
M_{11} & \propto & (A^2-{\rm const.})\,y^2,\nonumber\\
M_{12} & \propto & Ay,\\
M_{22} & \propto & A^2.\nonumber
\eea
This demonstrates that the generation of non-zero neutrino masses
requires non-vanishing values of $g_{ij}$, $A$, $\ye$ and/or $\yz$.
In the opposite limit of vanishing $A$, $\ye$, $\yz$ etc.\ and small
values of $\lambda_D''$, one finds
\bea
M_{11} & = & 0,\nonumber\\
M_{12} & = & 0, \label{eq:4.6}\\
M_{22} & \propto & \lambda_D'',\hspace*{20mm}\nonumber
\eea
i.e.\ the neutrino masses are proportional to the doublet scalar mass
splitting.

The diagonalization of the neutrino mass matrix $D_\nu=U^T_\nu M_\nu
U_\nu=(0,m_{\nu_2},m_{\nu_3})$ with the PMNS matrix $U_\nu$ leads to two
non-zero Majorana neutrino masses,
while the third mass is always zero. The observed neutrino mass
differences \cite{Patrignani:2016xqp} then translate directly to
absolute neutrino masses. A third non-zero neutrino mass could in
principle be accomodated, but would require at least an additional
scalar or fermion singlet. Assuming normal mass hierarchy, the
experimental constraints on the neutrino masses and mixing angles
translate directly into constraints on the couplings \cite{Casas:2001sr}
\beq
 g = U_M D_M^{-\frac{1}{2}} R D_{\nu}^{\frac{1}{2}}U_{\nu}^T,
 \label{eq:4.7}
\eeq
where $U_M$ diagonalises $M$ via $D_M=U_M^TMU_M$ and the rotation matrix
\beq
 R = 
 \begin{pmatrix}
  0 & \cos \varphi & - \sin \varphi \\
  0 & \sin \varphi &   \cos \varphi
 \end{pmatrix}
\eeq
depends on a single parameter $\varphi$. For definiteness, we take
all parameters entering Eq.\ (\ref{eq:4.7}) to be real, {\it i.e.}
the Dirac and Majorana phases in $U_\nu$ and the phase associated
with $R$ are assumed to be zero.

From Eq.\ (\ref{neutmass}) it is clear that in the absence of
all scalar-fermion couplings $g_{ij}$, neutrino masses cannot be
generated. The neutrino masses depend, however, also on the dark
mass spectrum and, through the fermion and scalar mixing matrices
$U_F$ and $U_S$, on the Yukawa couplings $y_1$ and $y_2$, the
singlet-doublet scalar coupling $A$ and, to a lesser extent, on
the exclusively singlet or doublet scalar couplings $\lambda_S$,
$\lambda_D$ etc. This can also be seen from Fig.\ 2 of Ref.\
\cite{Restrepo:2013aga} before electroweak symmetry breaking,
which involves all four dark mass parameters as well as the
couplings $g_{ij}^2$, $y_1$ or $y_2$, and $A$, but not $\lambda_S$,
$\lambda_D$ etc. In the following numerical studies of the total
neutrino mass sum, we therefore fix the latter to values of
${\cal O}(10^{-1})$ or smaller. 

\begin{figure}
\begin{center}
 \includegraphics[width=0.95\textwidth]{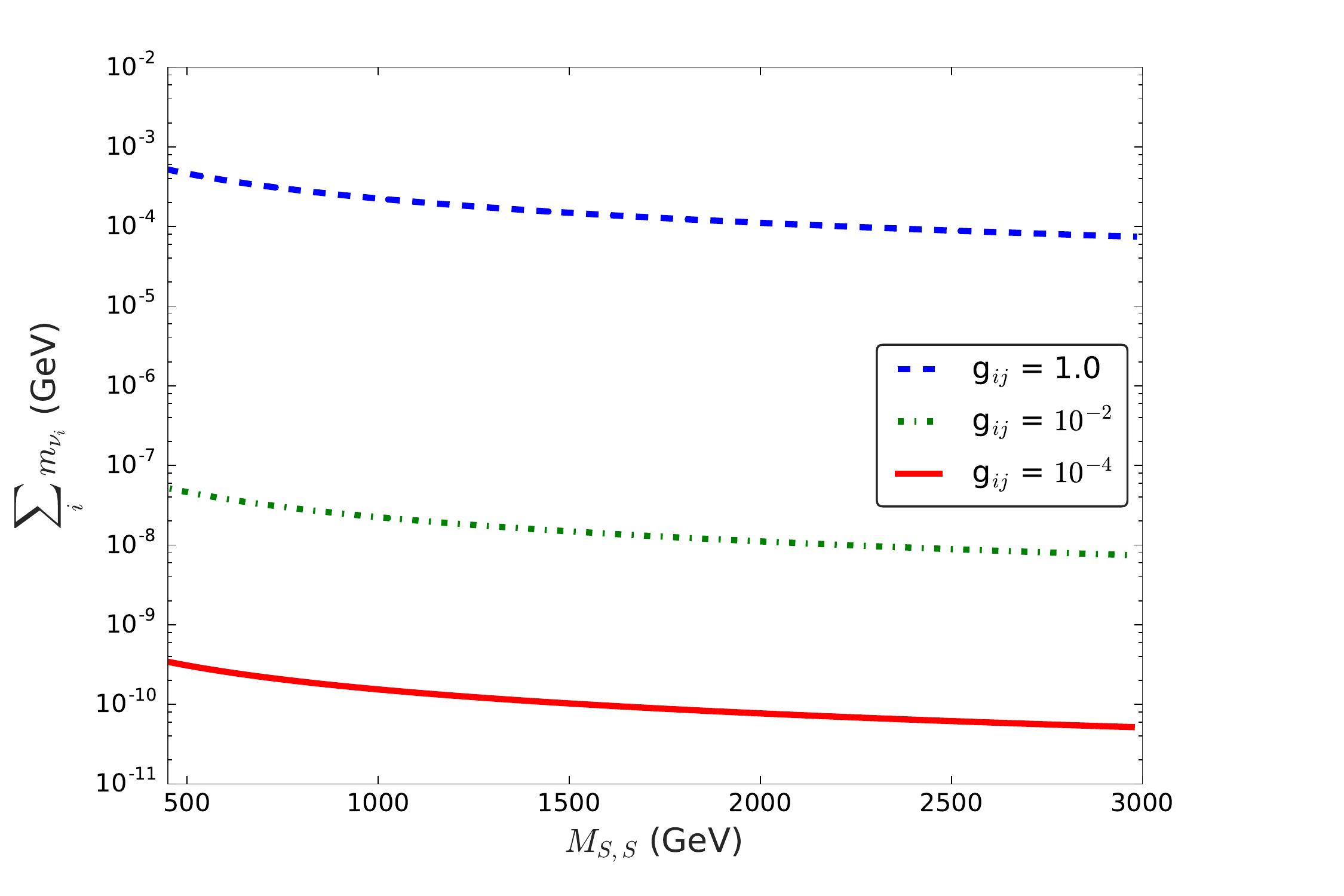}
\end{center}
\caption{Influence of the scalar-fermion couplings $g_{ij}$ on the sum of
 neutrino masses as a function of the singlet scalar mass parameter $\mss$.}
\label{fig:7}
\end{figure}

In Fig.\ \ref{fig:7}, we demonstrate the influence of the
scalar-fermion couplings $g_{ij}$ on the sum of neutrino masses
as a function of the scalar singlet mass $\mss$. The other masses
are scaled to it via $\msd=1.5\,\mss$, $\mfs=2\,\mss$ and $\mfd=
2.5\,\mss$, and the scalar and Yukawa couplings are small
($A=10^{-2}$ GeV, $y_1=2\cdot10^{-2}$, and $y_2=10^{-1}$).
Therefore, the singlet scalar and doublet fermion dominate, at
least to some extent, in the loop. As the singlet scalar mass
(and with it all other masses) increases, the neutrino masses
decrease moderately as expected from the propagator suppression
in the loop. More importantly, as the scalar-fermion couplings
$g_{ij}$, and in particular the singlet scalar-doublet fermion
couplings $g_{1i}$, decrease from values of 1 (dashed blue line)
to 10$^{-2}$ (dot-dashed green line) and
10$^{-4}$ (full red line), the neutrino mass sum drops over
eight orders of magnitude in agreement with the quadratic
scaling of $M_{\nu,ij}$ with $g_{ij}$ in Eq.\ (\ref{neutmass}).
In this scenario, the viable region of $\sum_i m_{\nu_i}=
{\cal O}(10^{-11})$ GeV is thus reached only for very small
values of $g_{ij}$.

\begin{figure}
\begin{center}
\includegraphics[width=0.95\textwidth]{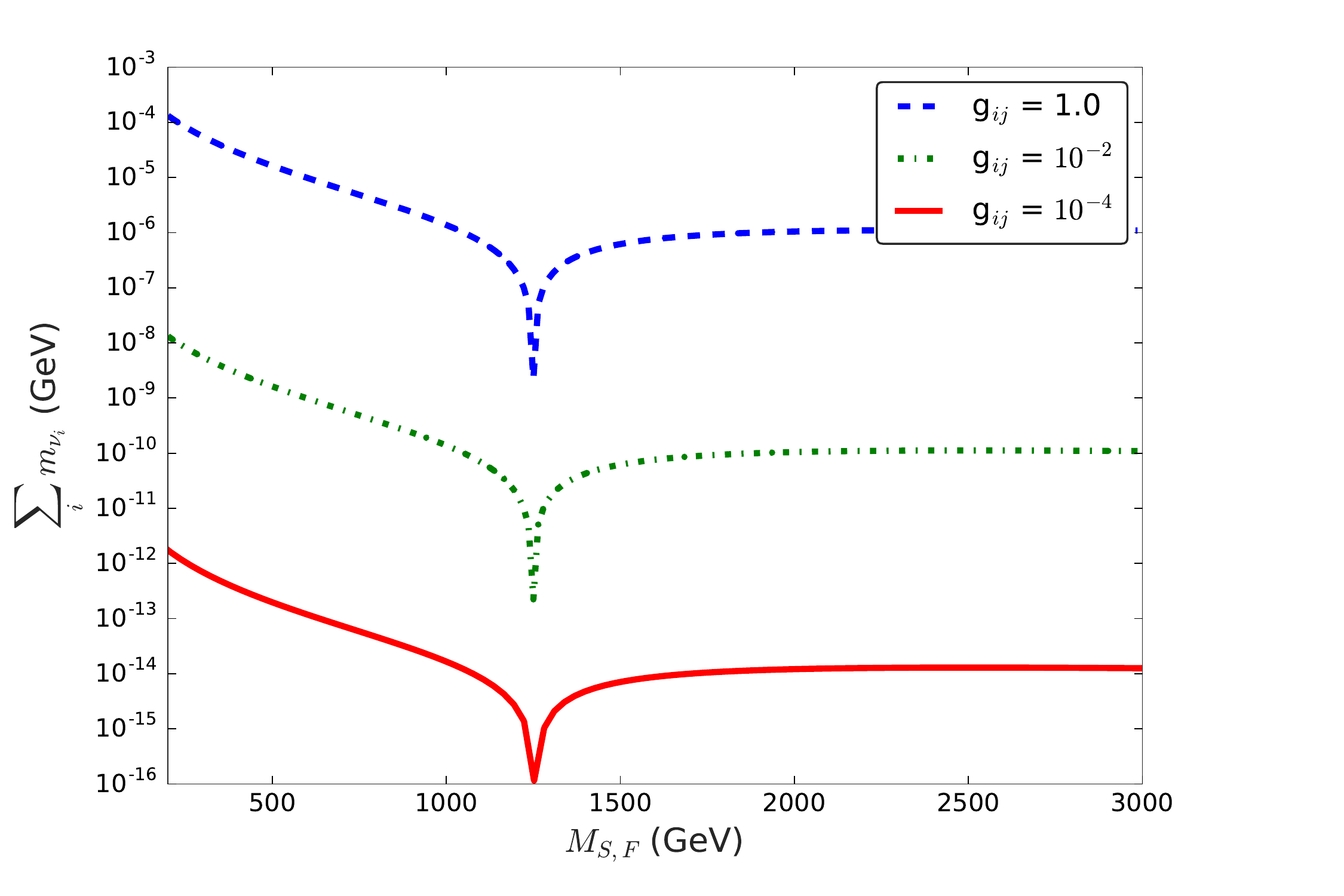}
\end{center}
\caption{Influence of the scalar-fermion couplings $g_{ij}$ on the sum of
 neutrino masses as a function of the singlet fermion mass parameter $\mfs$.}
\label{fig:8}
\end{figure}

This is, however, not always the case, as we demonstrate in Fig.\
\ref{fig:8}. Here, the neutrino mass sum is shown as a function of
the singlet fermion mass $\mfs$, while the other masses are scaled
to it via $\mss=2.5\,\mfs$, $\msd=3.5\,\mfs$ and $\mfd=1.5\,\mfs$.
The doublet scalar mass is substantially larger than in the previous
figure, and it mixes strongly ($A=10$ GeV) with the singlet scalar,
while the Yukawa couplings are comparable ($y_1=4\cdot10^{-2}$,
$y_2=2\cdot 10^{-2}$), so that the dominance of the singlet fermion
and the doublet scalar in the loop is less pronounced. In total, this
leads to neutrino mass sums that are at least an order of magnitude
smaller and fall more steeply for small $\mfs$. The striking feature
in Fig.\ \ref{fig:8} is the cancellation of terms of opposite sign
in Eq.\ (\ref{neutmass}) that leads to a vanishingly small neutrino
mass sum around $\mfs=1.25$ TeV. Outside this region, one observes
the same quadratic scaling with $g_{ij}$ from 1 (dashed blue line)
over $10^{-2}$ (dot-dashed green line) to $10^{-4}$ (full red line)
as before, but due to the larger masses and cancellations, the viable
neutrino mass region can now already be reached for intermediate
values of $g_{ij}$.

\begin{figure}[th]
\begin{center}
\includegraphics[width=0.95\textwidth]{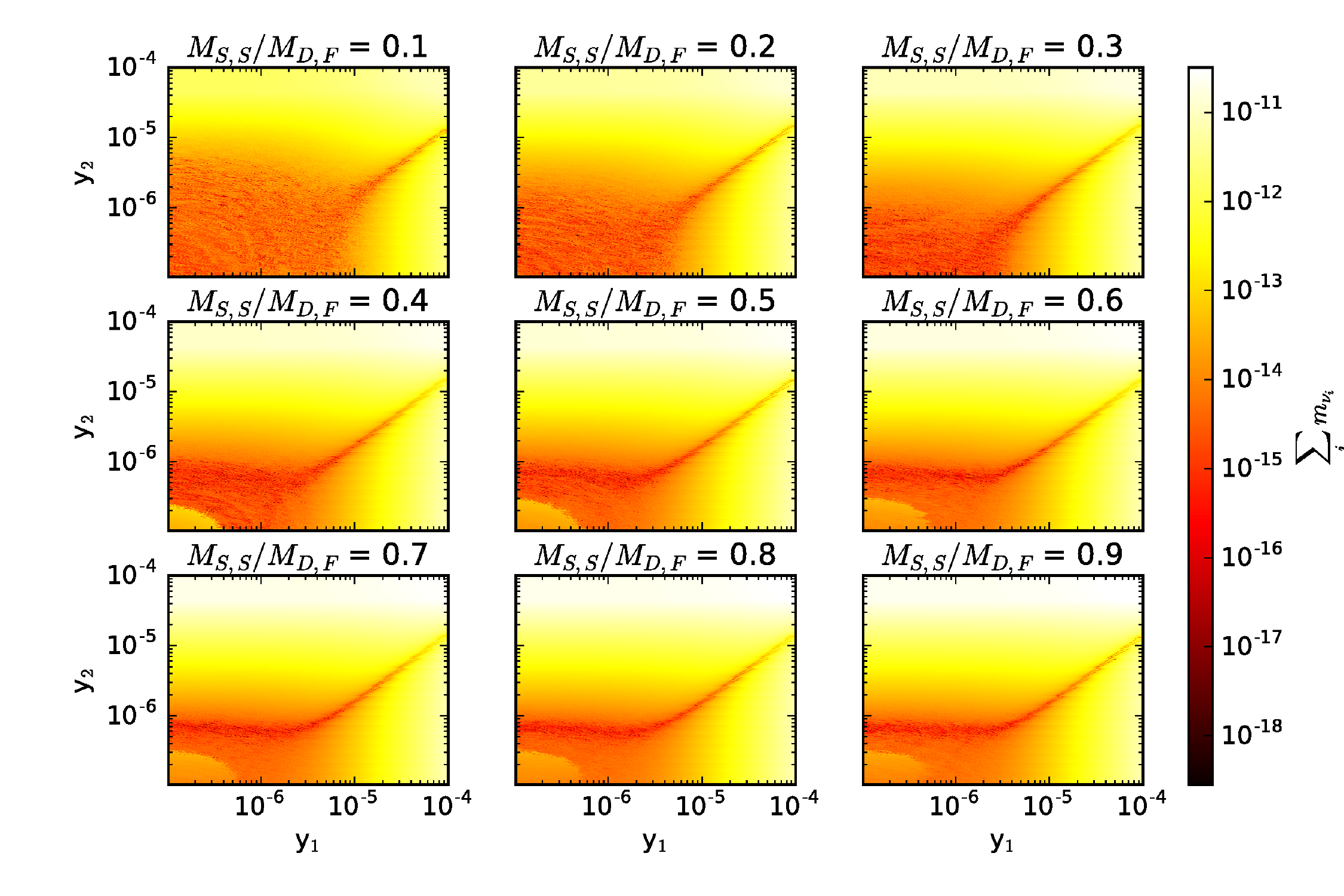}
\end{center}
\caption{Influence on the ratio of the singlet scalar mass $\mss=1$ TeV 
 over the singlet fermion mass $\mfs$ on the sum of neutrino masses in the
 plane of Yukawa couplings $y_1$--$y_2$. The doublet scalar and singlet fermion
 have been decoupled via $\mss/\msd=0.6$ and $\mss/\mfs=0.3$.}
\label{fig:9}
\end{figure}

In Fig.\ \ref{fig:9}, we investigate the influence of the Yukawa
couplings $y_1$ and $y_2$ on the sum of neutrino masses for
singlet scalar dark matter of mass $\mss=1$ TeV, which does
not mix with the doublet scalar ($A=0$), but couples to the
doublet fermion with fixed strength $g_{1i}=0.1$. The other
couplings $\lambda_S$, $\lambda_D$ etc.\ and $g_{2i}$ have no
significant impact and have been set to values of $10^{-5}$.
The doublet scalar and singlet fermion have thus also no
obvious direct influence, and their masses have been decoupled
with fixed mass ratios of $\mss/\msd=0.3$ and $\mss/\mfs=0.6$,
respectively. For large doublet fermion masses (top left), one
observes that the full range of Yukawa couplings shown leads
to a sum of neutrino masses of ${\cal O}(10^{-11})$ GeV or
smaller. If both Yukawa couplings $y_1$ and $y_2$ are of similar
size, they can be larger than when this is true for only one
of them. As the doublet fermion mass decreases and approaches
the scalar singlet mass (bottom right), the Yukawa couplings
also drop in order for the neutrino masses to remain in the
viable observed range. Interestingly, they can not be too small
either once the doublet fermion mass drops below the singlet
fermion mass (beyond $\mss/\mfs=0.3$). The doublet fermion then
has to mix more with the singlet fermion to reduce the neutrino
mass to the viable region.

\section{Lepton flavour violation}
\label{sec:5}

\begin{figure}
  \centering
  \begin{minipage}{\linewidth}
    \centering
    \begin{minipage}{0.25\textwidth}
      \begin{picture}(310,100)(0,0)
	\put(5,5){\mbox{\resizebox{!}{0.9\textwidth}{\includegraphics{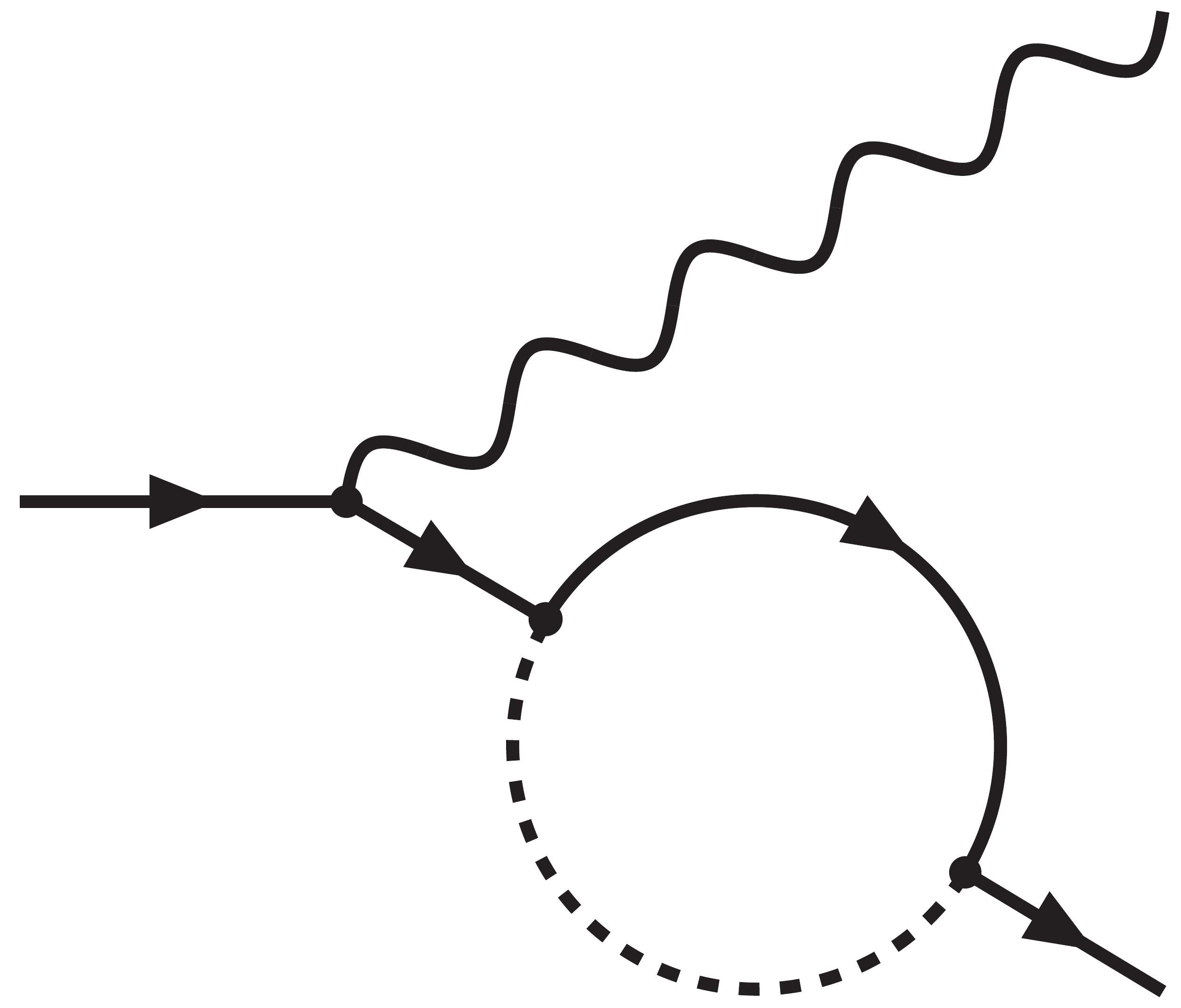}}}}
	\put( -5,  45){$e^-_k$}
	\put(120,  105){$\gamma$}
	\put(120, 5){$e^-_m$}
	\put( 85,  60){$\psi^-$ ($\chi_i$)} 
	\put( 15,  10){$X_i$ ($\phi^-$)}
      \end{picture}
    \end{minipage}
    \hspace{1.5cm}
    \begin{minipage}{0.25\textwidth}
      \begin{picture}(310,100)(0,0)
	\put(5,5){\mbox{\resizebox{!}{0.9\textwidth}{\includegraphics{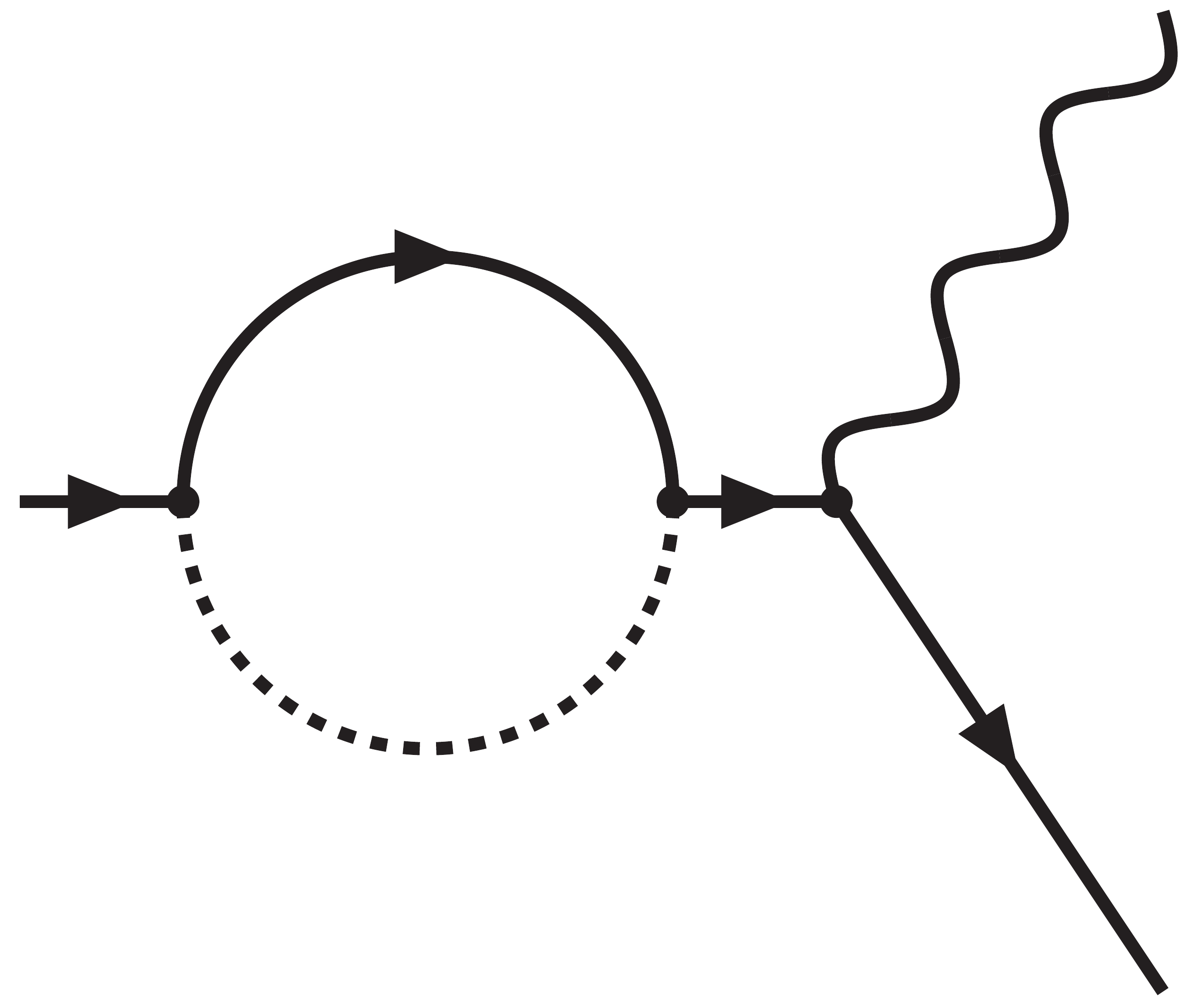}}}}
	\put( -5,  45){$e^-_k$}
	\put(120,  105){$\gamma$}
	\put(120, 5){$e^-_m$}
	\put( 25,  85){$\psi^-$ ($\chi_i$)}
	\put( 25,  15){$X_i$ ($\phi^-$)}
      \end{picture}
    \end{minipage}
  \end{minipage}
  \begin{minipage}{\linewidth}
\vspace*{8mm}
    \centering
    \begin{minipage}{0.25\textwidth}
      \begin{picture}(310,100)(0,0)
	\put(5,5){\mbox{\resizebox{!}{0.9\textwidth}{\includegraphics{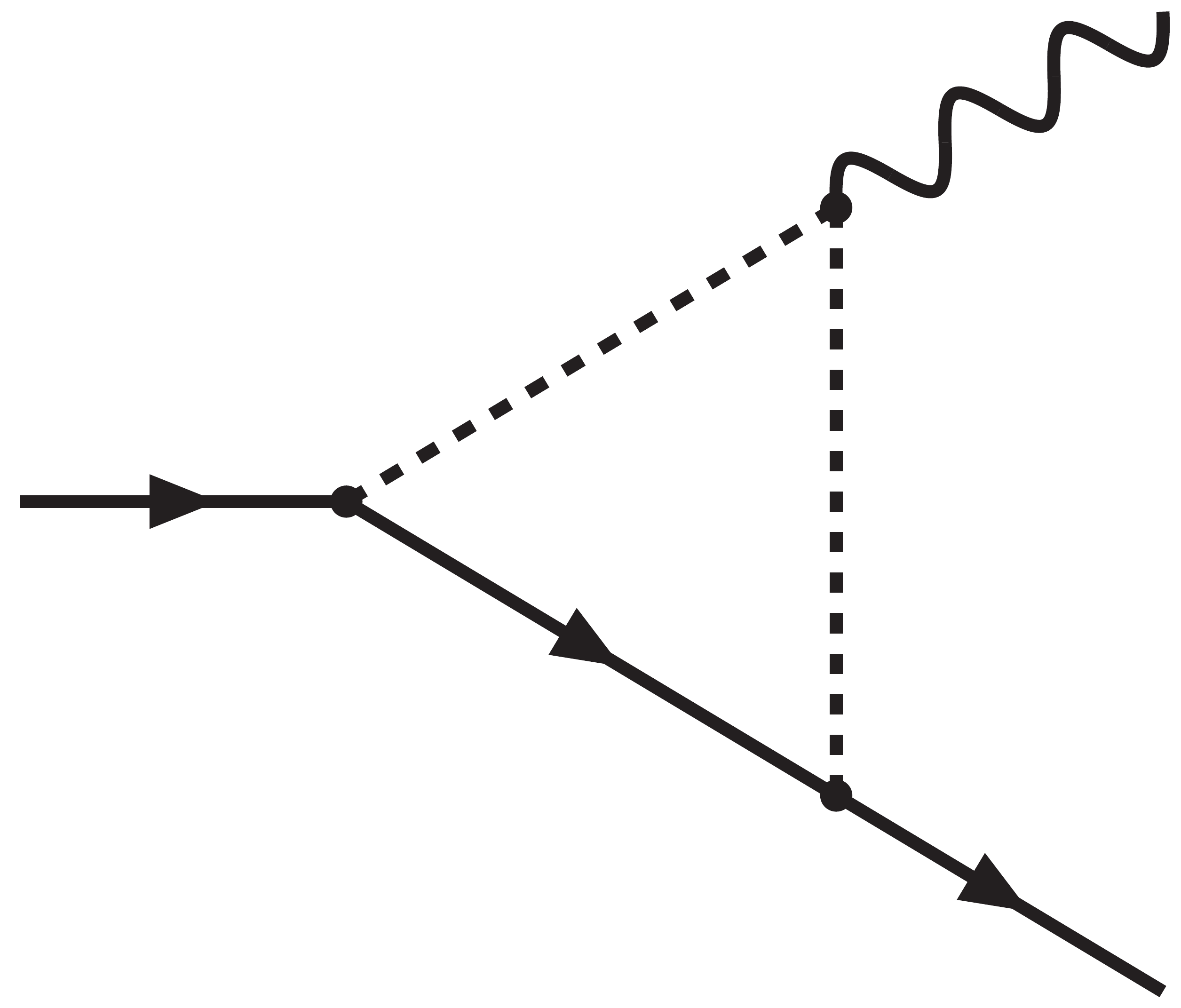}}}}
	\put( -5,  45){$e^-_k$}
	\put(120,  105){$\gamma$}
	\put(120, 5){$e^-_m$}
	\put( 50,  75){$\phi^-$ } 
	\put( 90,  45){$\phi^-$ } 
	\put( 50,  25){$\chi_i$}
      \end{picture}
    \end{minipage}
    \hspace{1cm}
    \begin{minipage}{0.25\textwidth}
      \begin{picture}(310,100)(0,0)
	\put(5,5){\mbox{\resizebox{!}{0.9\textwidth}{\includegraphics{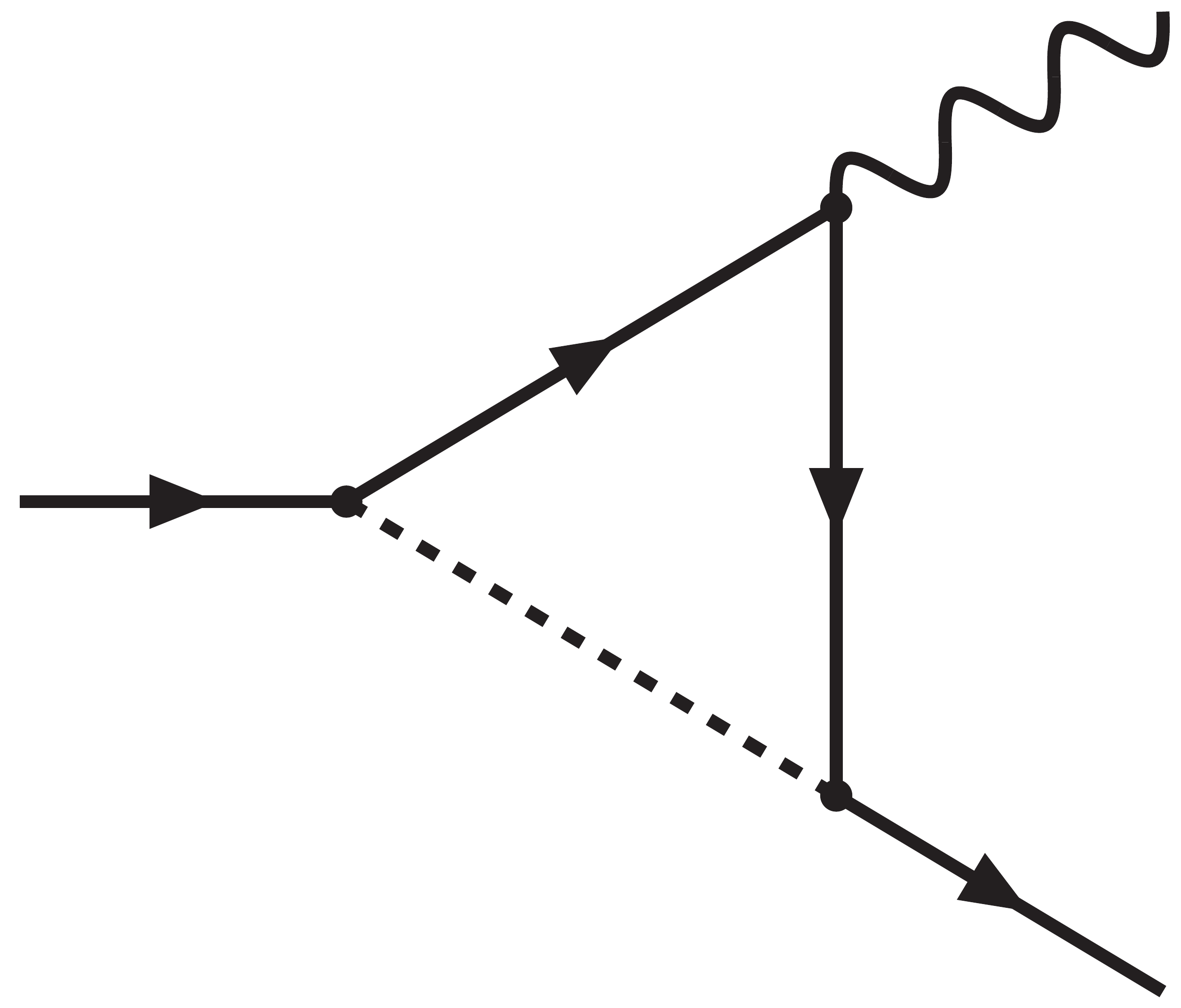}}}}
	\put( -5,  45){$e^-_k$}
	\put(120,  105){$\gamma$}
	\put(120, 5){$e^-_m$}
	\put( 50,  75){$\psi^-$ } 
	\put( 90,  45){$\psi^-$ } 
	\put( 50,  22){$X_i$}     
      \end{picture}
    \end{minipage}
  \end{minipage}
  \caption{Radiative one-loop processes $e_k^-\to e_m^-\gamma$ violating lepton flavour.}
  \label{fig:10}
\end{figure}
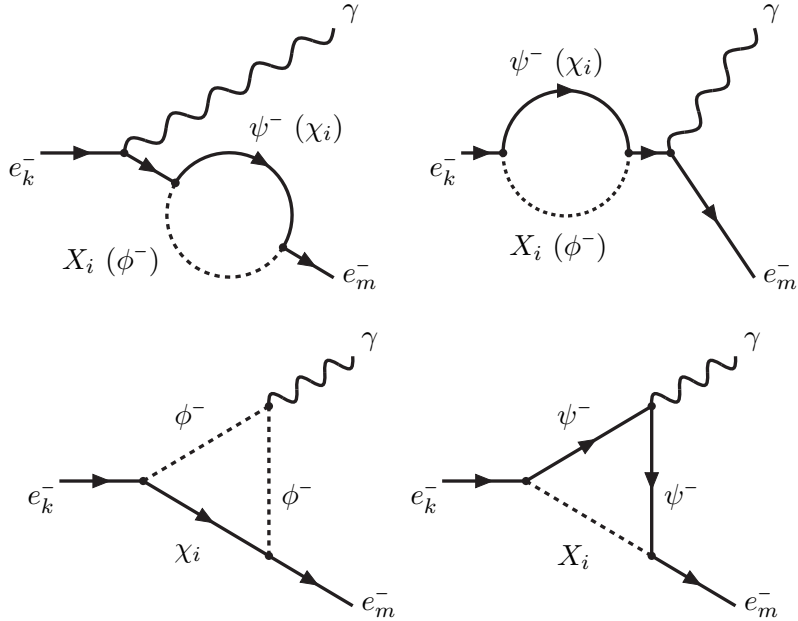

\begin{figure}
\begin{center}
\begin{minipage}{0.3\textwidth}\begin{picture}(310,100)(0,0)
      \put(5,5){\mbox{\resizebox{!}{4cm}{\includegraphics{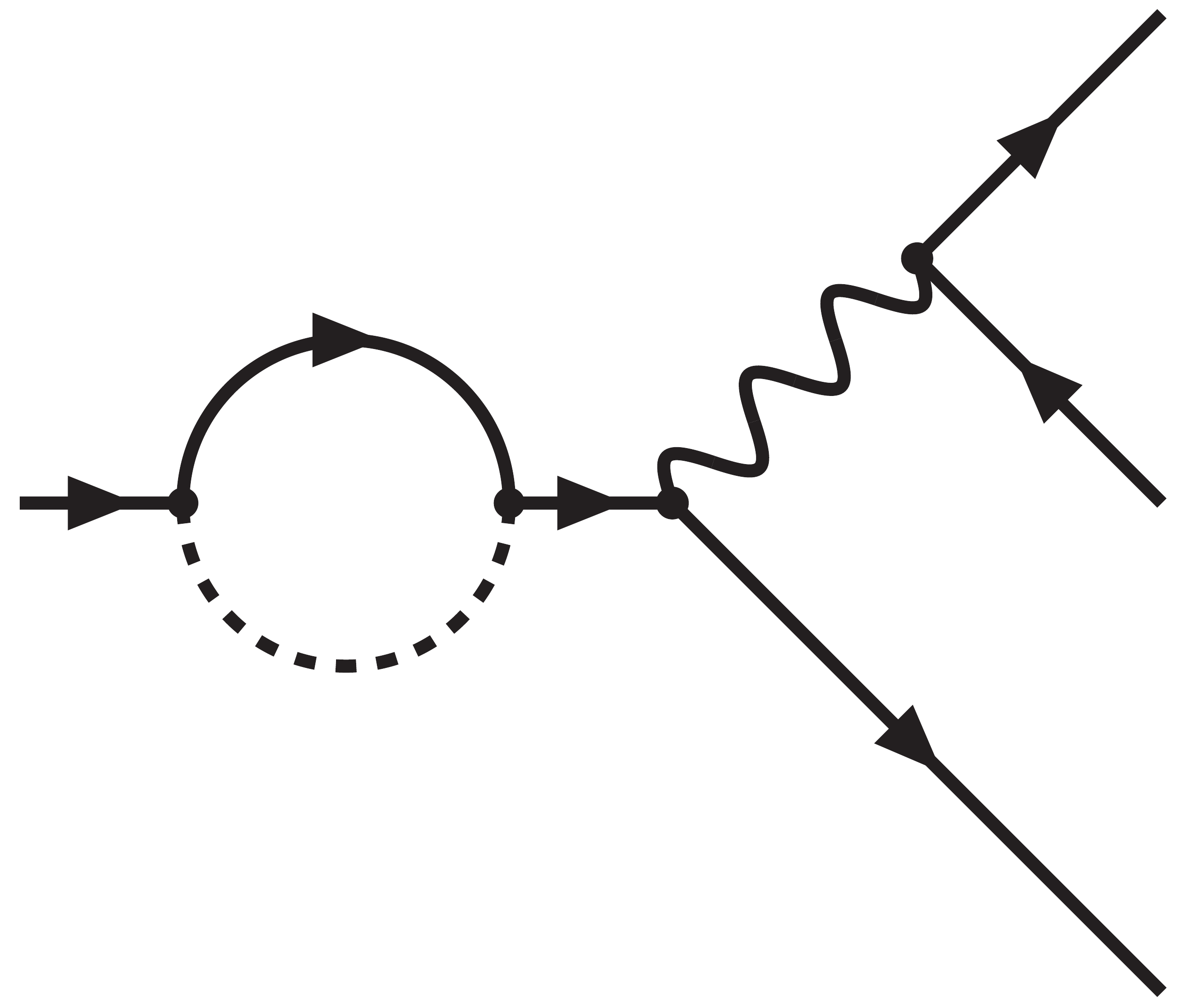}}}}
      \put( 10,  70){$e^-_k$}
      \put( 67,  70){$e^-_n$}
      \put( 30,  90){$\chi_i$ ($\psi^-$)} 
      \put( 30,  30){$\phi^-$ ($X_i$)}
      \put( 85,  85){$\gamma$}
      \put(110, 110){$e^-_m$}
      \put(110,  60){$e^+_m$}
      \put( 95,  25){$e^-_n$}
\end{picture}\end{minipage}
\hspace{5mm}
\begin{minipage}{0.3\textwidth}\begin{picture}(310,100)(0,0)
      \put(5,5){\mbox{\resizebox{!}{4cm}{\includegraphics{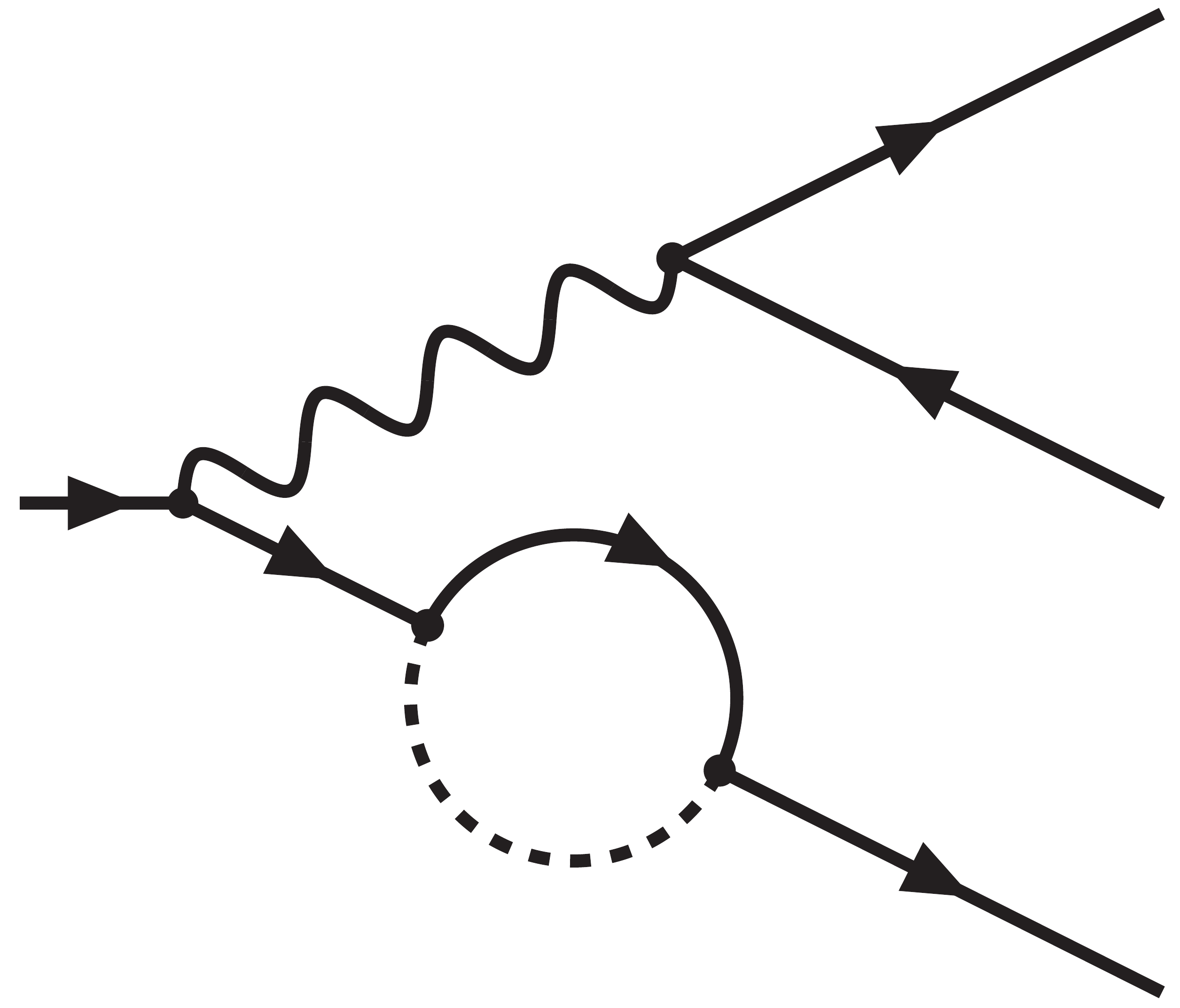}}}}
      \put( 10,  70){$e^-_k$}
      \put( 30,  40){$e^-_k$}
      \put( 60,  65){$\chi_i$ ($\psi^-$)} 
      \put( 60,  10){$\phi^-$ ($X_i$)}
      \put( 50,  90){$\gamma$}
      \put(105, 115){$e^-_m$}
      \put(105,  60){$e^+_m$}
      \put(105,  5){$e^-_n$}
\end{picture}\end{minipage}
\hspace{5mm}
\begin{minipage}{0.3\textwidth}\begin{picture}(310,100)(0,0)
      \put(5,5){\mbox{\resizebox{!}{4cm}{\includegraphics{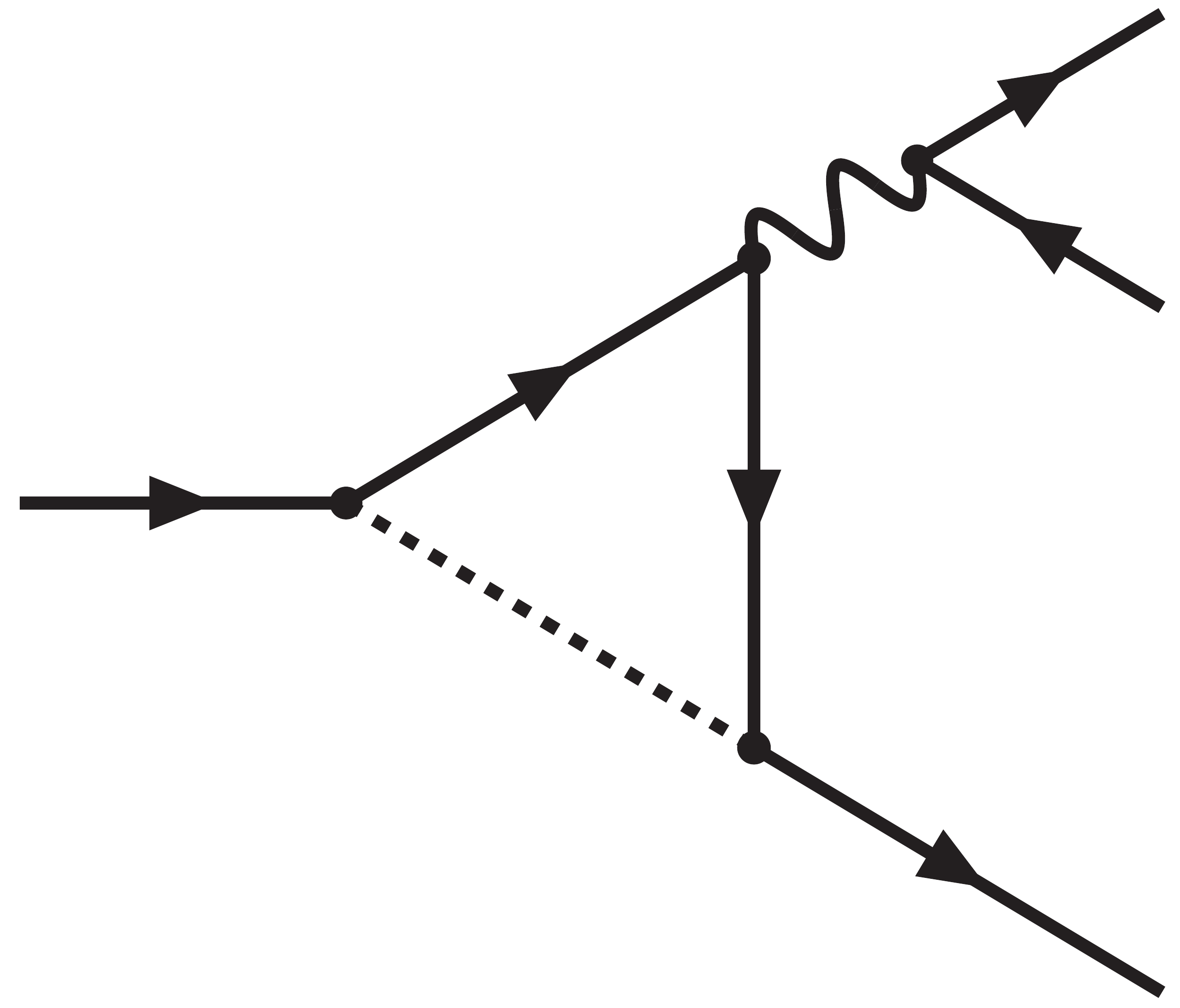}}}}
      \put( -5,  60){$e^-_k$}
      \put( 95,  105){$\gamma$}
      \put( 60,  85){$\psi^-$} 
      \put( 60,  30){$X_i$}
      \put( 98,  60){$\psi^-$}
      \put(115, 115){$e^-_m$}
      \put(115, 80){$e^+_m$}
      \put(115,  2){$e^-_n$}
\end{picture}\end{minipage}
\\[15mm]
\hspace{4mm}
\begin{minipage}{0.3\textwidth}\begin{picture}(310,100)(0,0)
      \put(5,5){\mbox{\resizebox{!}{4cm}{\includegraphics{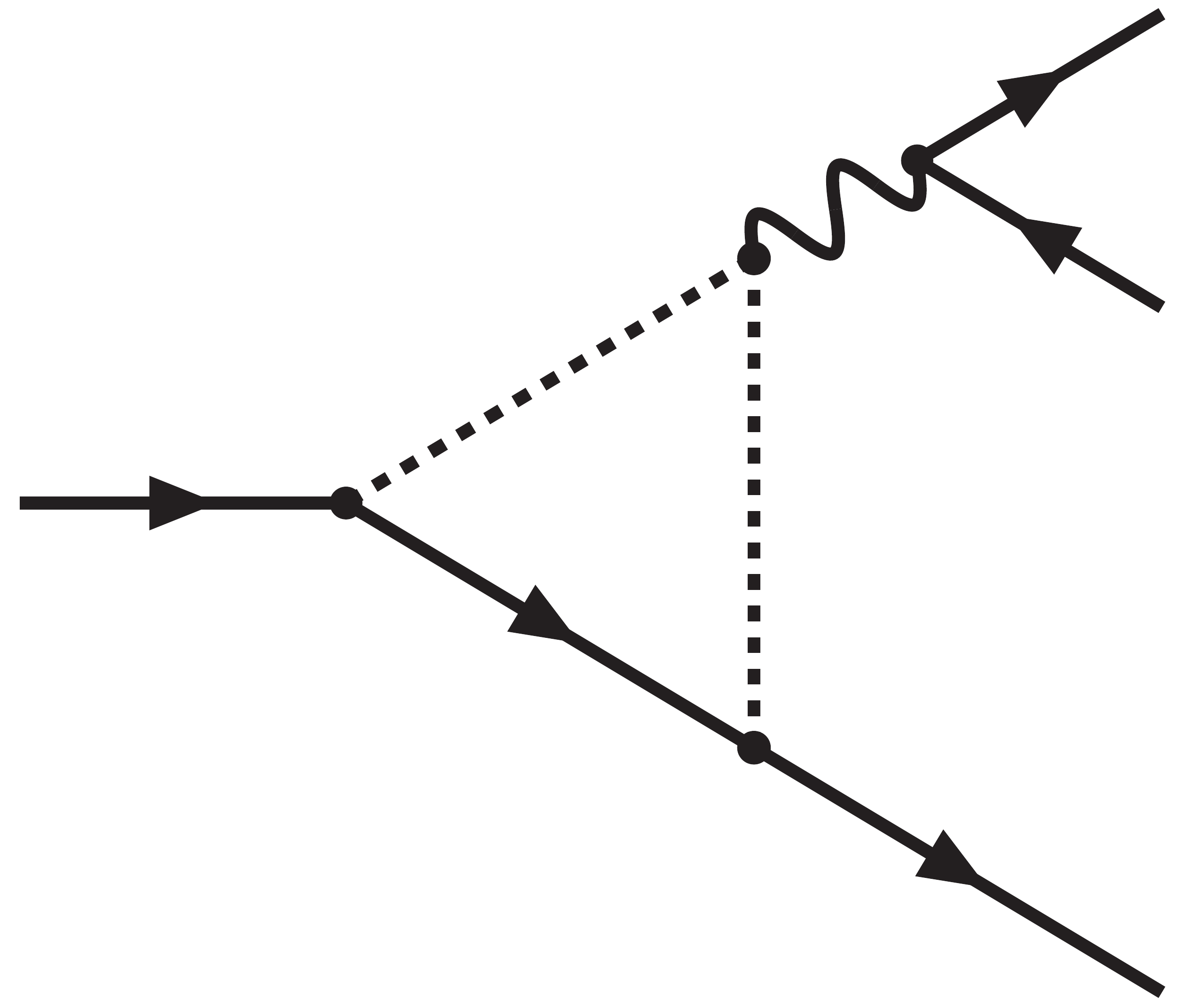}}}}
      \put( -5,  60){$e^-_k$}
      \put( 95,  105){$\gamma$}
      \put( 60,  85){$\phi^-$} 
      \put( 60,  30){$\chi_i$}
      \put( 98,  60){$\phi^-$}
      \put(115, 115){$e^-_m$}
      \put(115, 80){$e^+_m$}
      \put(115,  2){$e^-_n$}
\end{picture}\end{minipage}
\hspace{9mm}
\begin{minipage}{0.3\textwidth}\begin{picture}(310,100)(0,0)
      \put(5,5){\mbox{\resizebox{!}{4cm}{\includegraphics{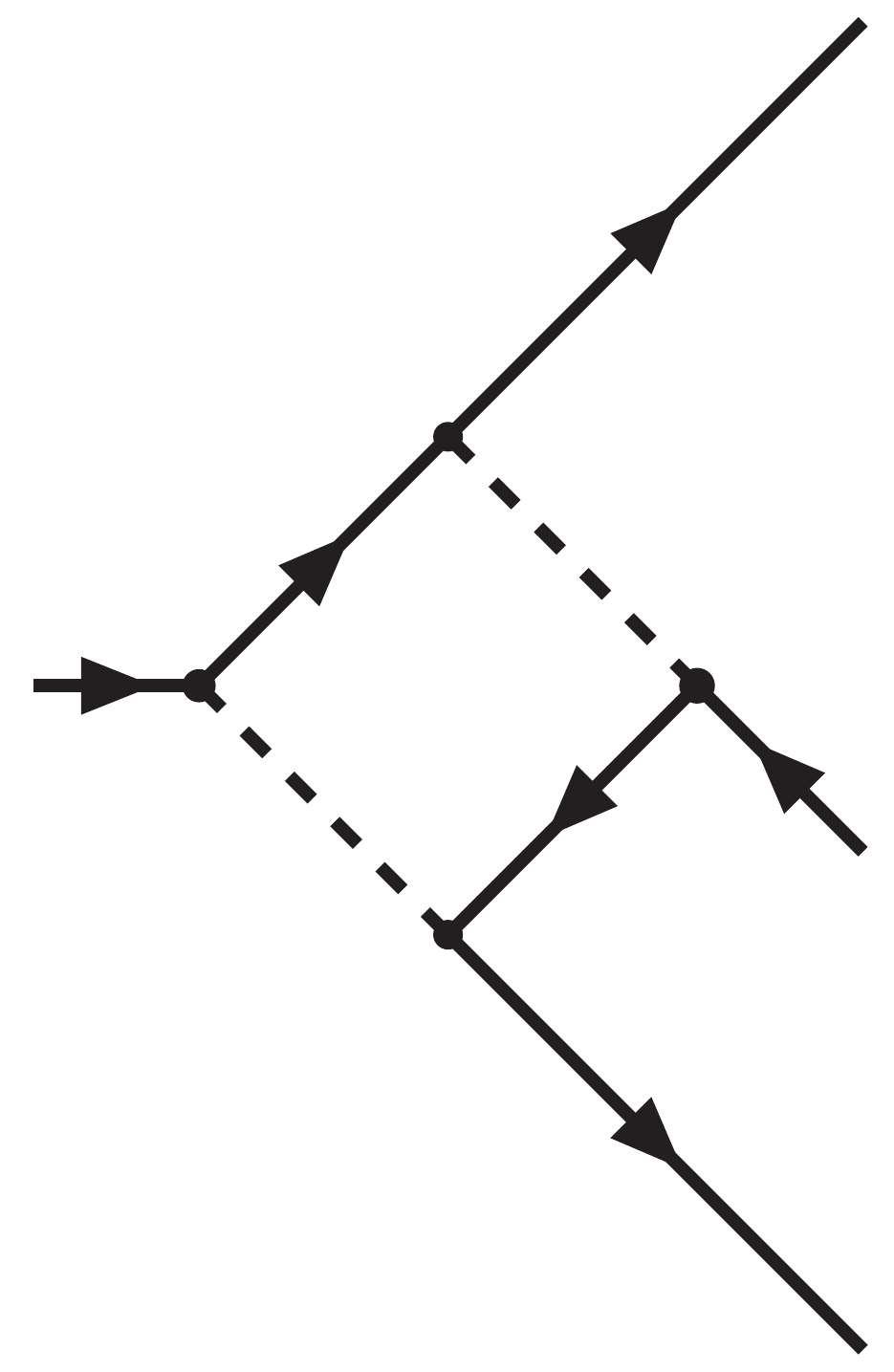}}}}
      \put( -5,  60){$e^-_k$}
      \put( 60,  75){$\phi^-$ ($X_j$)}
      \put( -2,  80){$\chi_i$ ($\psi^-$)} 
      \put( -2,  42){$\phi^-$ ($X_i$)}
      \put( 50,  40){$\chi_j$ ($\psi^-$)}
      \put(80, 115){$e^-_m$}
      \put(75,  55){$e^+_n$}
      \put( 80,  2){$e^-_o$}
\end{picture}\end{minipage}
\hspace{-4mm}
\begin{minipage}{0.3\textwidth}\begin{picture}(310,100)(0,0)
      \put(5,5){\mbox{\resizebox{!}{4cm}{\includegraphics{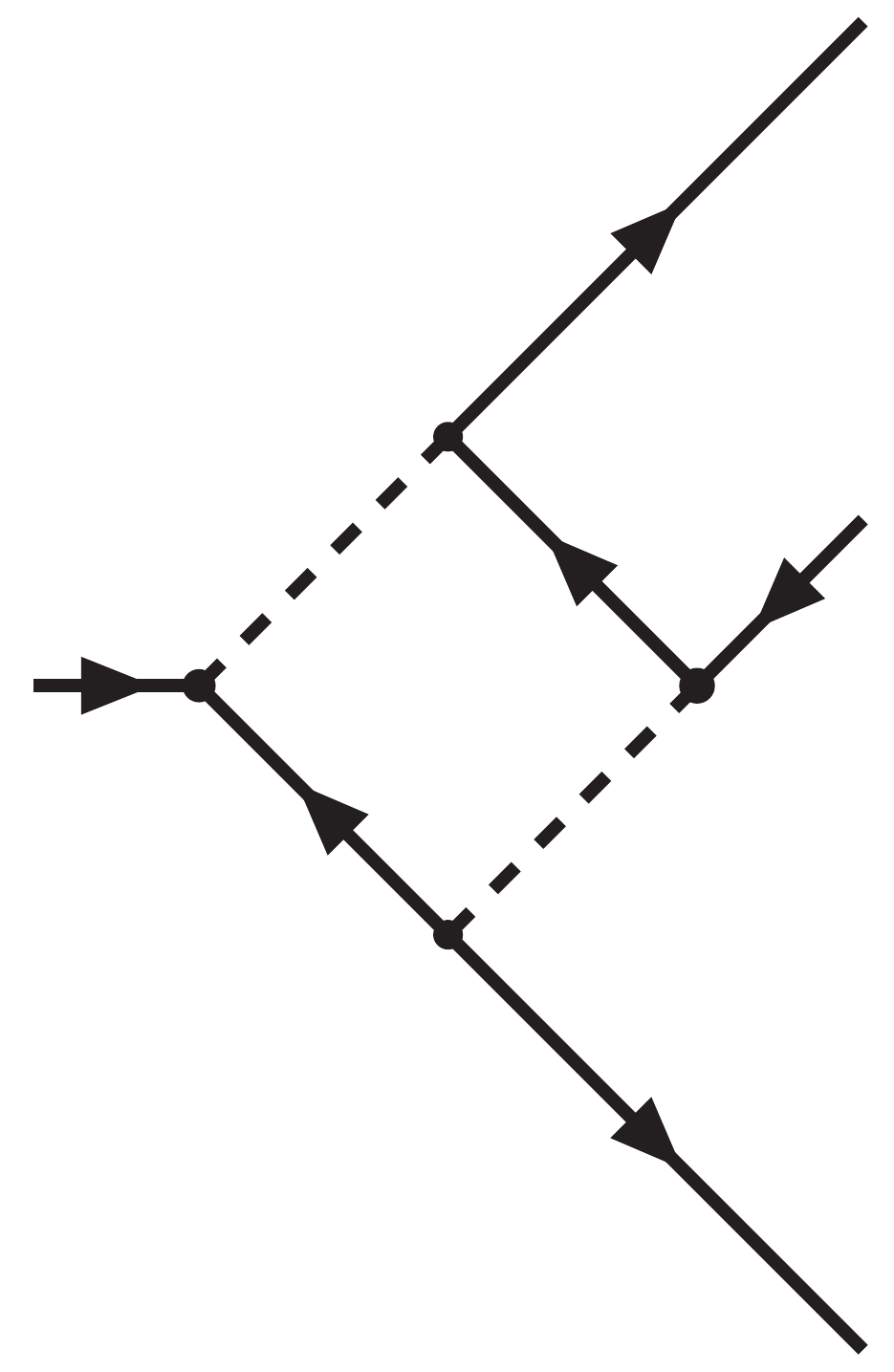}}}}
      \put( -5,  60){$e^-_k$}
      \put( 55,  80){$\psi^-$ ($\chi_j$)}
      \put( -2,  80){$X_i$ ($\phi^-$)} 
      \put( -2,  42){$\psi^-$ ($\chi_i$)}
      \put( 55,  45){$X_j$ ($\phi^-$)}
      \put(80, 115){$e^-_m$}
      \put(75,  60){$e^+_n$}
      \put( 80,  2){$e^-_o$}
\end{picture}\end{minipage}
\end{center}
\caption{Loop diagrams contributing to the leptonic decays $e_k^- \rightarrow e_m^- e_n^+ e_o^-$}
\label{fig:11}
\end{figure}
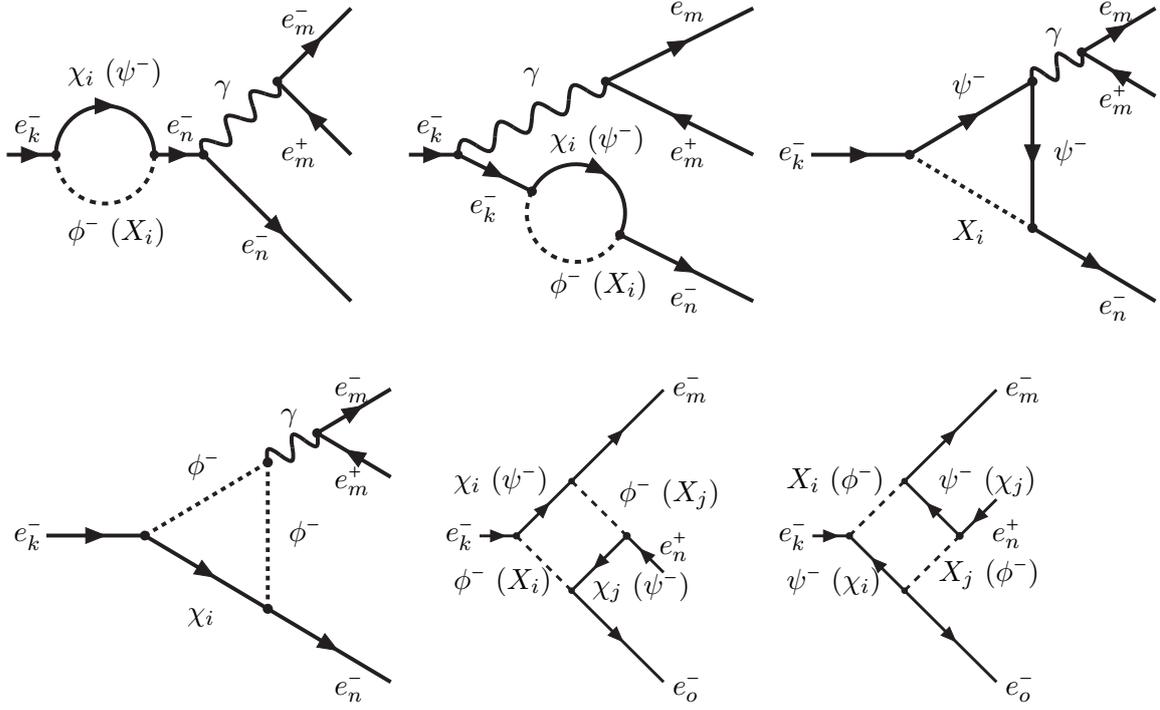

The radiative generation of neutrino masses in our model implies the
existence of lepton-number violating terms in the Lagrangian and thus
of lepton-flavour violating processes. Important examples are the
radiative transitions $\mu\to e\gamma$ etc., the leptonic decays
$\mu\to3e$ etc., and conversion processes such as $\mu$\,Au$\to e$\,Au.
The first class of transitions arises through the bubble (top) or
triangle (bottom) diagrams depicted in Fig.\ \ref{fig:10}, where
always both new scalars and fermions must run in the loops.
Neglecting the external lepton masses, the branching ratio
can be expressed as
\bea
 {\rm BR}(\mu \rightarrow e \gamma) & = & \frac{3 \alpha_{em}}{64 \pi G_F^2}
 \left[ \frac{1}{2 m_{\psi^-}^4} \left( \sum_i g_{11} g_{12} U_{S,1i}^2 F(\frac{m_{X_i}^2}{m_{\psi^-}^2}) \right)^2
 + \frac{1}{m_{\phi^-}^4} \left( \sum_i g_{21} g_{22} U_{F,1i}^2 F(\frac{m_{\chi_i}^2}{m_{\phi^-}^2} )\right)^2 \right. \nonumber \\
 && \left. \hspace{1.2cm}
 + \frac{1}{m_{\phi^-}^2m_{\psi^-}^2} \left( \sum_i g_{11} g_{12} U_{S,1i}^2  F(\frac{m_{X_i}^2}{m_{\psi^-}^2}) \right) \left( \sum_j g_{21} g_{22} U_{F,1j}^2 F(\frac{m_{\chi_j}^2}{m_{\phi^-}^2})\right) \right]
 \label{brMuToEG}
\eea
with 
\beq
 F(x) = \frac{ 2 x^3 + 3 x^2 - 6 x^2 \ln x - 6 x + 1}{6(x-1)^4}
\eeq
and similarly for $\tau\to e\gamma$ and $\tau\to\mu\gamma$.
It is obvious that in the limit of vanishing scalar-fermion
couplings $g_{ij}\to0$, not only the neutrino masses, but also
the lepton-flavour violating processes disappear. While the
neutrino mass matrix in Eq.\ (\ref{neutmass}) always depends
on both the dark scalar and fermion mixing matrices $U_S$ and
$U_F$, the charged lepton processes always require one charged
dark particle in the loop, so that they depend at the amplitude
level only on one of the neutral mixing matrices.
The same diagrams as in Fig.\ \ref{fig:10} also contribute to
conversion processes, when the photon is taken off-shell and
couples to the heavy nucleus. 
The flavour-violating leptonic decay processes $\mu\to3e$ etc.\
are mediated by the diagrams shown in Fig.\ \ref{fig:11}.
They involve bubble, triangle and box one-loop topologies and thus
at the amplitude level two (bubbles and triangles) or four (boxes)
powers of scalar-fermion couplings $g_{ij}$ and mixing matrices.

\begin{figure}
\begin{center}
 \includegraphics[width=0.95\textwidth]{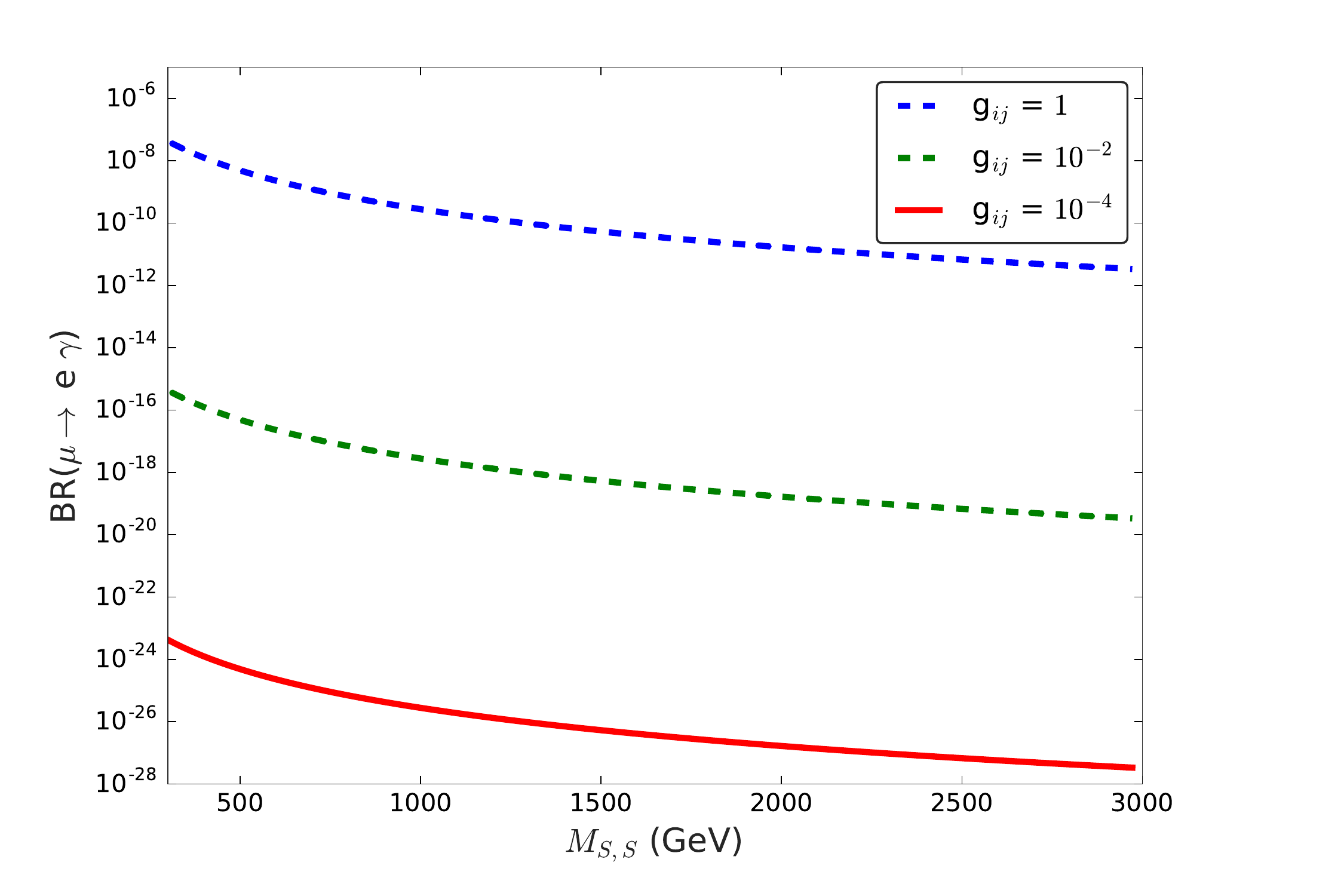}
\end{center}
\caption{Influence of the scalar-fermion couplings $g_{ij}$ on the
 branching ratio BR$(\mu\to e\gamma)$ as a function of the singlet
 scalar mass parameter $\mss$.}
\label{fig:12}
\end{figure}

Current and future limits on the branching ratios for the processes
described above are listed in Tab.\ \ref{lfvlimits}.
\begin{table}
\centering
\caption{Current and future limits on lepton-flavour violating processes.}
\label{lfvlimits}
 \begin{tabular}{l|l l|l l}
 Process & \multicolumn{2}{c}{ Current limit } & \multicolumn{2}{c}{ Future expectation }   \\
 \hline
 \hline
 $\mu  \rightarrow e \gamma$   			& 5.7$\times 10^{-13}$ 	& \cite{Adam:2013mnn}		& 6$\times 10^{-14}$			& \cite{Baldini:2013ke}	\\
 $\tau \rightarrow e \gamma$   			& 3.3$\times 10^{-8}$  	& \cite{Aubert:2009ag}		& $\approx$  3$\times 10^{-9}$		& \cite{Aushev:2010bq}	\\
 $\tau \rightarrow \mu \gamma$   		& 4.4$\times 10^{-8}$  	& \cite{Aubert:2009ag}		& $\approx$  3$\times 10^{-9}$		& \cite{Aushev:2010bq}	\\
 $\mu  \rightarrow 3 e$   			& 1.0$\times 10^{-12}$ 	& \cite{Bellgardt:1987du}		& $\approx$  $ 10^{-16}$		&\cite{Blondel:2013ia}		\\
 $\tau \rightarrow 3 \mu$   			& 2.1$\times 10^{-8}$  	& \cite{Hayasaka:2005xw}		& $\approx$  $ 10^{-9}$			& \cite{Aushev:2010bq}	\\
 $\tau^{-} \rightarrow e^{-} \mu^{+}\mu^{-}$   	& 2.7$\times 10^{-8}$  	& \cite{Hayasaka:2005xw}		& $\approx$  $ 10^{-9}$			& \cite{Aushev:2010bq}	\\
 $\tau^{-} \rightarrow \mu^{-} e^{+}e^{-}$   	& 1.8$\times 10^{-8}$  	& \cite{Hayasaka:2005xw}		& $\approx$  $ 10^{-9}$			& \cite{Aushev:2010bq}	\\
 $\tau \rightarrow 3 e $   			& 2.7$\times 10^{-8}$  	& \cite{Hayasaka:2005xw} 		& $\approx$  $ 10^{-9}$			& \cite{Aushev:2010bq}	\\
 $\mu^{-}$\,Ti $\rightarrow e^{-}$\,Ti   		& 4.3$\times 10^{-12}$ 	& \cite{Dohmen:1993mp} 		& $\approx$  $ 10^{-18}$		& \cite{Sato:2008zzm} 	\\
 $\mu^{-}$\,Au $\rightarrow e^{-}$\,Au   		& 7.0$\times 10^{-13}$ 	& \cite{Bertl:2006up} 		& --					&			\\
 $\mu^{-}$\,Al $\rightarrow e^{-}$\,Al   		& --			&				& $10^{-15} - 10^{-18}$			& \cite{Litchfield:2014qea} \\
 $\mu^{-}$\,SiC $\rightarrow e^{-}$\,SiC 	  	& --			&				& $10^{-14}$				& \cite{Natori:2014yba} 	
\end{tabular}
\end{table}
As one can see, the process $\mu\to e\gamma$ typically sets
the most stringent limits on new particle masses and couplings.
We therefore show in Fig.\ \ref{fig:12} the dependence of the
branching ratio for this process on the singlet scalar mass
$\mss$. All other parameters have been set as in Fig.\
\ref{fig:7} to allow for an easy comparison with the
corresponding neutrino masses. In particular, we show again
three curves for scalar-fermion couplings $g_{ij}$ of 1
(dashed blue curve), $10^{-2}$ (dot-dashed green curve) and
$10^{-4}$ (full red curve). As expected, the branching ratio
falls with the singlet scalar mass in the loop, but much
more directly with the scalar-fermion couplings $g_{ij}$. 
Current and future experimental limits can be evaded already
with values of $10^{-2}$, whereas the neutrino masses required
even smaller values of about $10^{-4}$ for these couplings.

\section{Numerical results}
\label{sec:6}

In this section, we present our main numerical results. We focus on the
case of scalar dark matter and perform two random numerical scans, the
first over the full parameter space, the second in a region where
coannihilation processes of singlet scalar and doublet fermion processes
become important. We impose all available experimental constraints, in
particular on the mass $M_H=125$ GeV and SM couplings of the Higgs boson
\cite{Patrignani:2016xqp}, the observed dark matter relic density
$\Omega^{\rm obs} h^2=0.1186\pm0.0031$ \cite{Ade:2013zuv}, and the neutrino
masses and mixing angles \cite{Gonzalez-Garcia:2014bfa}. Limits on the
direct detection cross section and lepton-flavour violating processes
(cf.\ Tab.\ \ref{lfvlimits}) are shown explicitly for the first scan
and imposed on the second scan. In both cases, we have obtained ${\cal O}
(10^4)$ viable points. Possible LHC constraints are discussed at the
end of this section.

\subsection{Random scan}

In our first random scan, we vary all new masses from 200 GeV to 2 TeV,
the scalar couplings in the range $|\lambda_S|,|\lambda_D|,|\lambda_D'|,
\lambda_D''\in[0;2\pi]$, the singlet-doublet scalar coupling $A\in[0;10^4]$
GeV, the Yukawa couplings $|y_1|,y_2\in[0;1]$, and the scalar-fermion
couplings are either zero or varied in the range $|g_{ij}|\in[0;2\pi]$.
They are then constrained directly by the neutrino masses and mixing
angles through the parameterisation in Eq.\ (\ref{eq:4.7}).

\begin{figure}
\begin{center}
\includegraphics[width=0.95\textwidth]{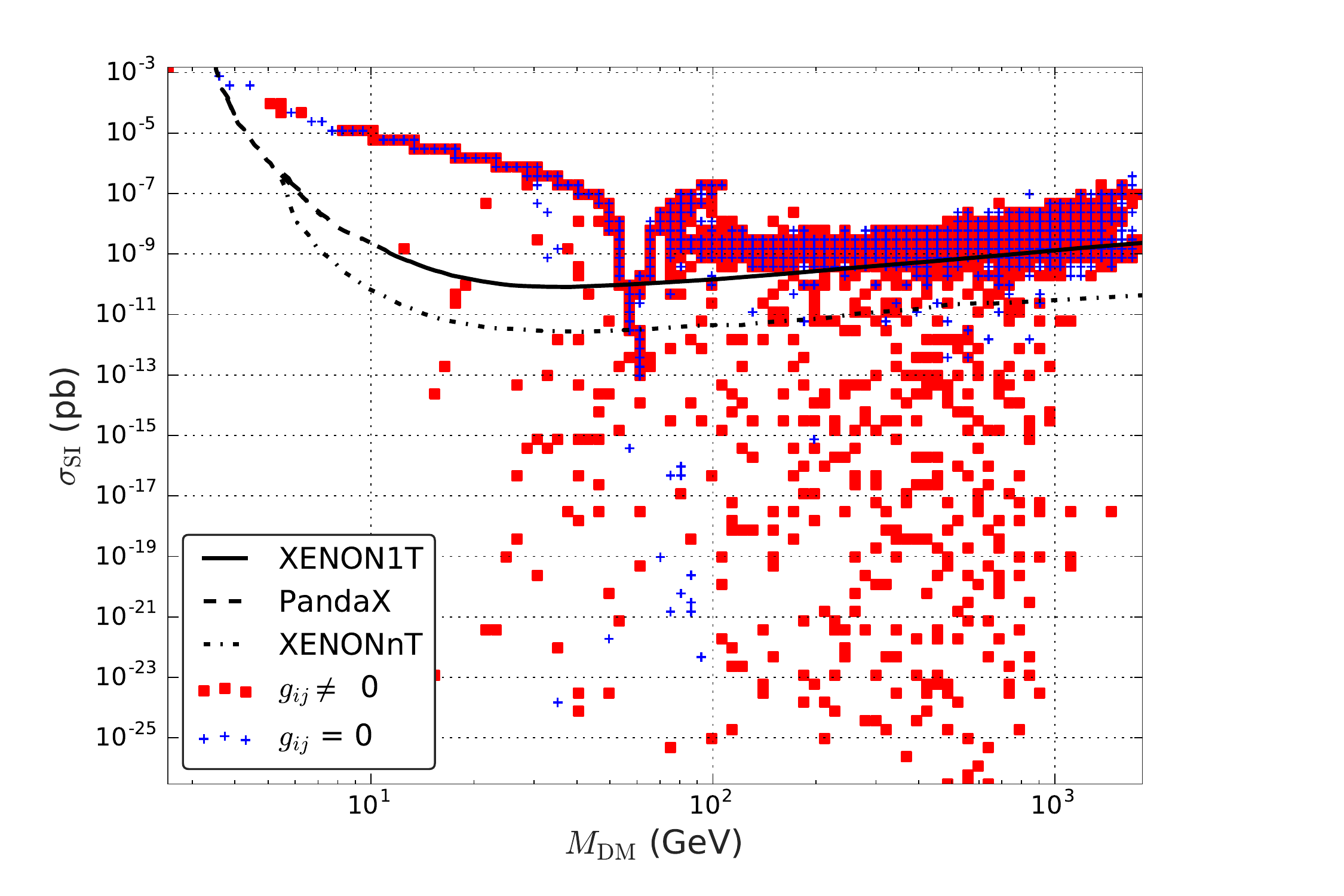}
\end{center}
\caption{Direct detection cross section of scalar singlet dark matter
 as a function of its mass without (blue) and with (red) coupling of
 the scalar to the fermion sector.}
\label{fig:13}
\end{figure}

In Fig.\ \ref{fig:13}, we plot the spin-independent direct detection
cross section $\sigma_{\rm SI}$ as a function of the physical mass of
scalar dark matter, which can be a mixture of the singlet and doublet.
In particular, their mixing can be large and lead to physical masses
of a few GeV despite the fact that the scanned mass parameters lie
above 200 GeV. When the scalar dark matter does not couple to
fermions (blue crosses), it is well known that the narrow relic
density constraint is strongly correlated with the direct
detection cross section. This leads to a narrow band below and
on the Higgs resonance at $M_{\rm DM}\simeq M_H/2$, which becomes
somewhat wider above. For constant $\Omega h^2$, the Higgs coupling
and direct detection cross section get smaller as the DM mass
increases towards the Higgs resonance. On the resonance,
annihilation is very efficient and the couplings and direct
detection cross section must be very small. As expected for
singlet-doublet scalar dark matter alone \cite{Cohen:2011ec,%
Cheung:2013dua}, most of the parameter points are now excluded
by PandaX \cite{Tan:2016diz,Liu:2017drf} (dashed curve) and
XENON1T \cite{Aprile:2017iyp} (full curve). Furthermore, they
obviously cannot explain the neutrino masses. The coupling
to the fermion sector opens up new regions of parameter space
(red squares) allowed by all experimental constraints that cannot even be
probed by XENONnT \cite{Aprile:2015uzo}, since the direct
detection cross section is independent of the couplings $g_{ij}$.
The correct relic density is reached through annihilation
processes into leptons and/or scalar-fermion coannihilations.

\begin{figure}
\begin{center}
\includegraphics[width=0.95\textwidth]{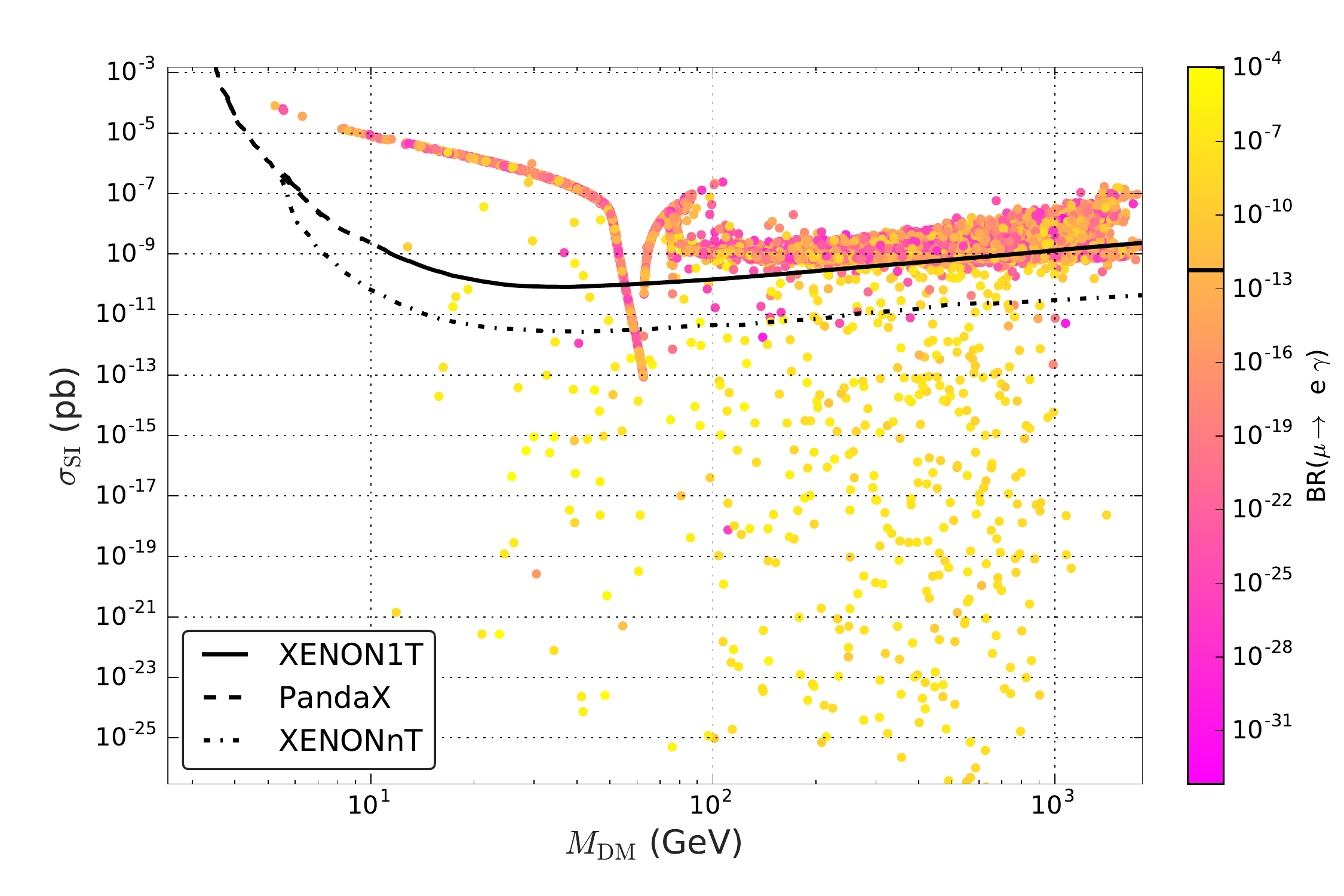}
\end{center}
\caption{Direct detection cross section of scalar singlet dark matter
 as a function of its mass and its correlation with the lepton-flavour
 violating branching ratio BR$(\mu\to e\gamma)$ (colours).}
\label{fig:14}
\end{figure}

As Fig.\ \ref{fig:14} shows, most of the new points are,
however, excluded by the lepton-flavour branching ratio
BR$(\mu\to e\gamma)$. Nevertheless, some of them fulfill
in fact all current experimental constraints. Interestingly,
they lead to direct detection cross sections that will soon
be tested by XENONnT.

\subsection{Coannihilation region}

In our second scan, we focus on scalar dark matter of mass
$\mss\in[10;3000]$ GeV, which can coannihilate with doublet
fermions of mass $\mfd=[1.05;1.2]\,\mss$, so that these
processes contribute at least 50\% to the relic density
cross section. Doublet scalars and singlet fermions have
larger masses of [$1.5\,\mss;3$ TeV] and [$\mfd;3$ TeV],
respectively. We limit the mixing in the scalar and fermion
sectors by reducing the scan ranges of $|\lambda_S|$, $A$,
$|y_1|$ and $y_2$ to values below $\pi\cdot10^{-4}$. This
also enhances the annihilation into lepton final states.

\begin{figure}
\begin{center}
\includegraphics[width=0.95\textwidth]{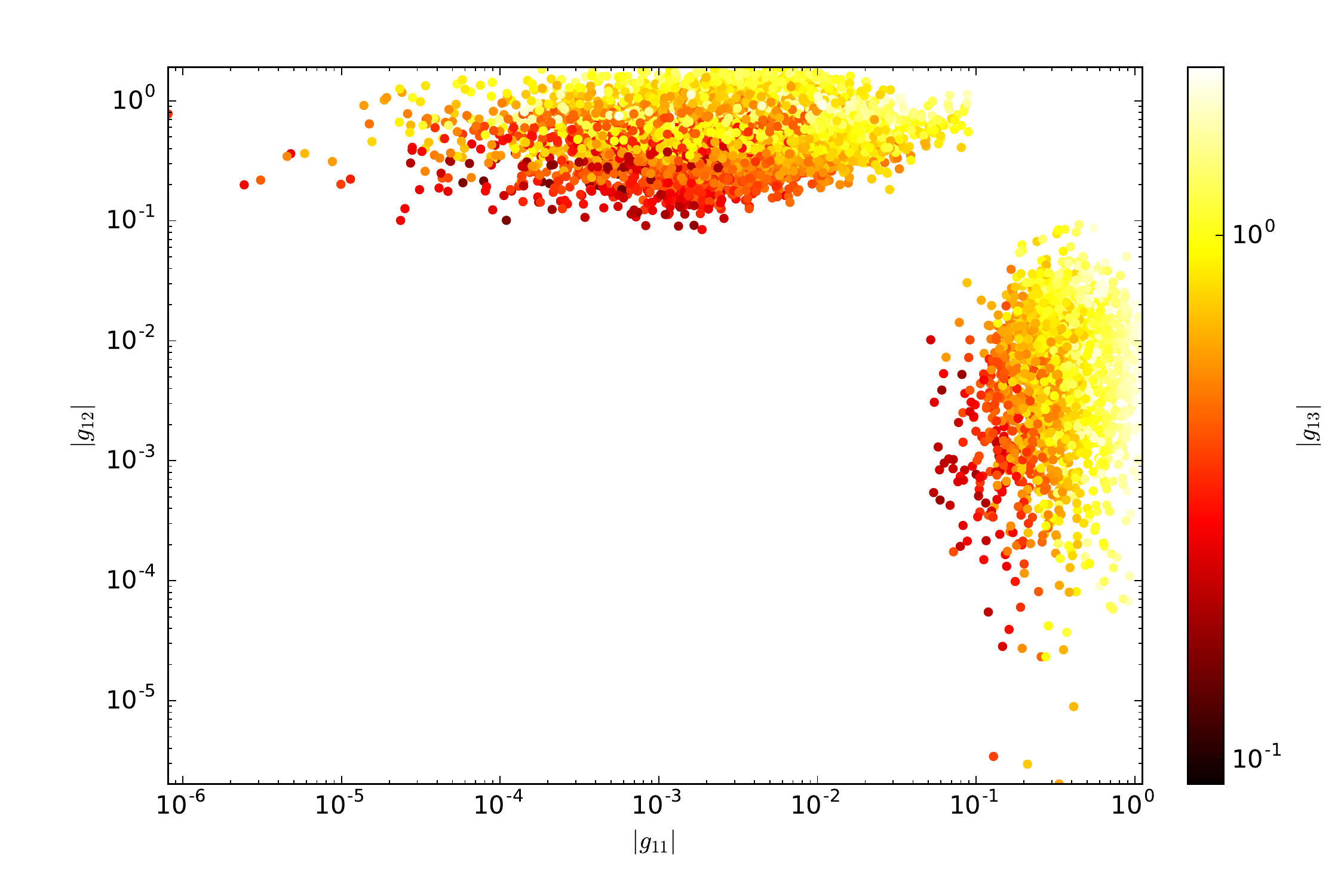}
\caption{Viable scalar-fermion couplings in the plane $|g_{11}|$--$|g_{12}|$
 (colour code for $|g_{13}|$).}
\label{fig:15}
\end{center}
\end{figure}

After imposing all experimental constraints, including those
on the direct detection cross section and lepton-flavour
violating processes, we obtain the scalar-fermion couplings
$g_{ij}$ shown in Fig.\ \ref{fig:15}. Since $A$, $y_1$ and
$y_2$ are now all small, we can obtain viable neutrino masses
for sizeable values of $g_{ij}$. As Fig.\ \ref{fig:15} shows,
at least one of these couplings must be large, but they cannot
be both large at the same time, which reflects the two rather
different neutrino mass differences.
Due to the weaker limits
on lepton-flavour violation for processes involving the $\tau$
lepton, the values of $g_{13}$ are less restricted than those
of $g_{11}$ and $g_{12}$. As the singlet scalar mass $\mss$ and
with it the doublet fermion mass $\mfd$ increase, so must the
couplings $g_{ij}$ to compensate for the propagator suppression
in the neutrino mass loops. Conversely, as $\lambda_D''$ and
with it the doublet scalar mass splitting and the neutrino
masses increase (cf.\ Eq.\ \ref{eq:4.6}), the corresponding
scalar-fermion couplings $g_{2i}$ must decrease for the neutrino
masses to remain in the viable range.

\begin{figure}[th]
\begin{center}
\includegraphics[width=0.95\textwidth]{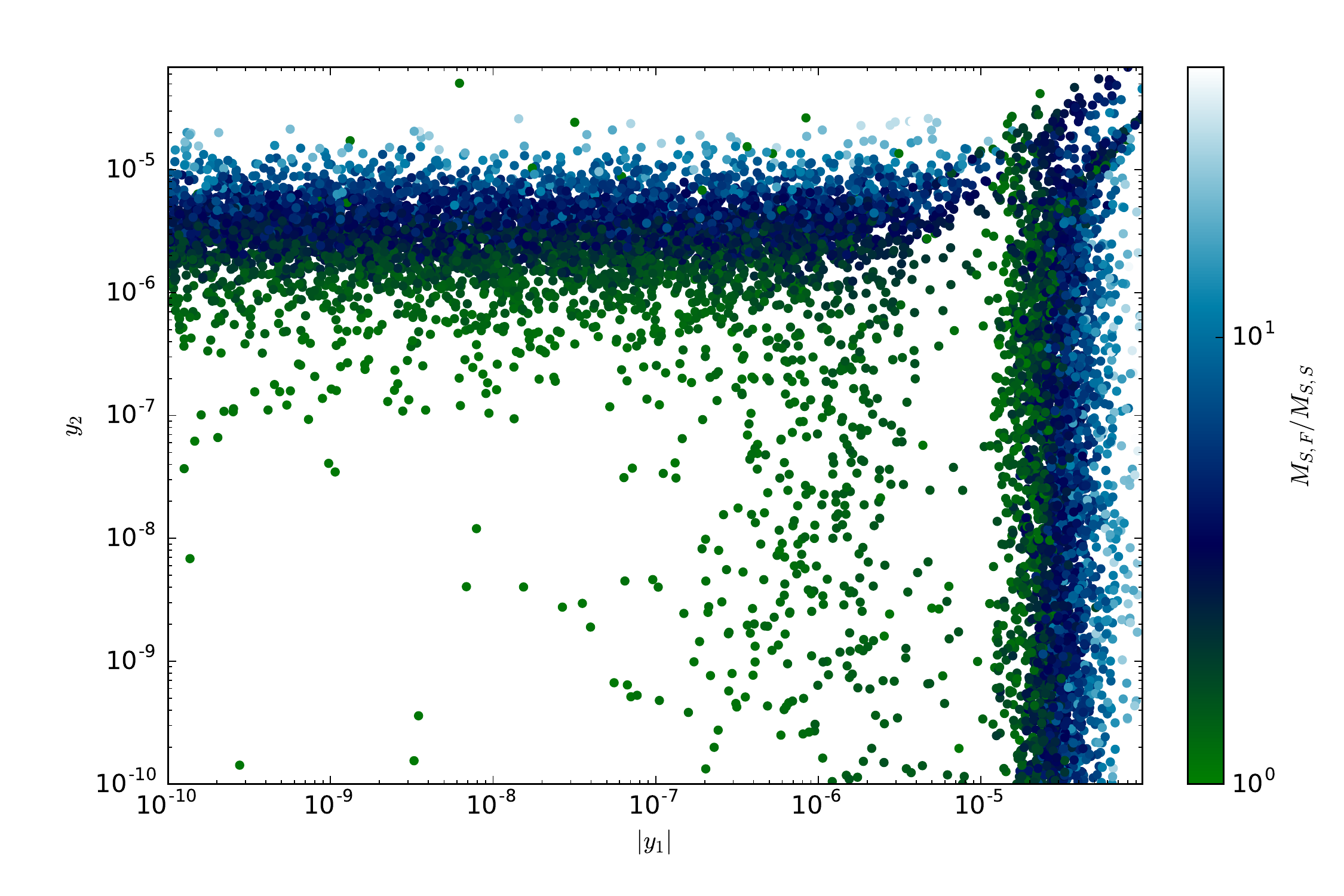}
\caption{Viable Yukawa couplings in the plane $|y_1|$--$|y_2|$
 (colour code for the mass ratio $\mfs/\mss$).}
\label{fig:16}
\end{center}
\end{figure}

A similar plot to the one in Fig.\ \ref{fig:15} for $g_{1i}$
shows the plane of the Yukawa couplings $|y_1|$--$y_2$ in Fig.\
\ref{fig:16}. Again, one of these couplings, but not
both must be relatively large, in particular when the singlet
fermion is decoupled (blue points). However, when it becomes
light (green points), mixes with the doublet fermion and
contributes to coannihilation, smaller values for both Yukawa
couplings also become viable.

\begin{figure}
\begin{center}
\includegraphics[width=0.95\textwidth]{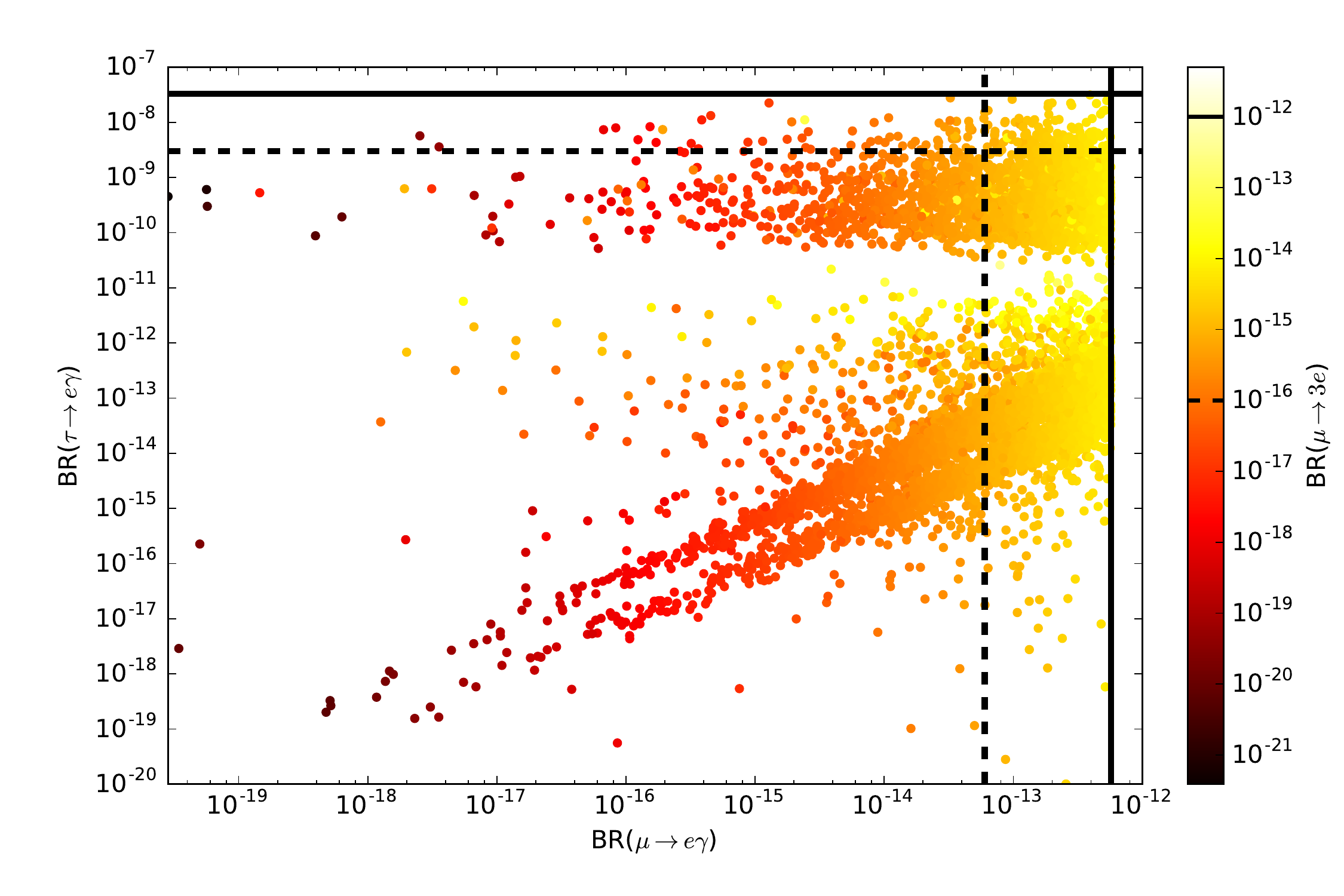}
\end{center}
\caption{Predicted branching ratios for the most sensitive lepton-flavour
 violating processes BR($\mu \rightarrow e \gamma$), BR($\tau \rightarrow
 e \gamma$) and  BR($\mu \rightarrow 3 e$) (colour) with current (full
 black lines) and future (dashed black lines) exclusion limits.}
\label{fig:17}
\end{figure}

The direct detection cross section
is governed by the coupling $\lambda_S$ of the scalar
singlet to the Higgs boson. It ranges from $10^{-27}$ pb
for $\lambda_S=10^{-9}$ to $10^{-14}$ pb for $\lambda_S=10^{-4}$
and is thus beyond the reach even of XENONnT. The situation
is therefore similar to the one in the previous section,
where many new viable models had very small direct detection
cross sections and had to be constrained by lepton-flavour
violating processes. We therefore present in Fig.\ \ref{fig:17}
the branching ratios BR($\mu\to e\gamma$), BR($\tau\to e\gamma$) and
BR($\mu\to 3e$). Current and future experimental limits are indicated
by full and dashed black lines. Since we impose the current limits
on our scan, no excluded models are found. A substantial fraction of
the viable models will be tested by future muon decay experiments
in either of the two channels, while the sensitivity of future tau
decay experiments remains limited and would have to be improved by
at least two orders of magnitude to completely probe the upper one
of the two viable parameter regions, where $g_{11}$ stays relatively
constant.

\subsection{LHC constraints}

In our first random scan over the full parameter space of scalar
dark matter, the latter can be a mixture of singlet and doublet.
Various collider limits on either case, in particular from
Higgs invisible decays at the LHC and the charged scalar
partners at LEP and the LHC, have been discussed in the past
\cite{Abe:2014gua}.

If dark matter is dominated by the singlet component, as it is
also the case in our second random scan over the scalar-fermion
coannihilation region, it couples neither to the photon nor to
the weak gauge bosons, but only to the Higgs boson
through the coupling $\lambda_S$. Constraints from the LHC thus
currently come only from the invisble decay width of the Higgs
boson in a mass region below 62.5 GeV. The upper limits lie currently
at 67\% for associated $ZH$ production in ATLAS \cite{Aaboud:2017bja}
and 24\% for a combination of different production channels in CMS
\cite{CMS:2018awd}. As our Fig.\ \ref{fig:14} and Fig.\ 8b in Ref.\
\cite{CMS:2018awd} show, models with such large couplings are
already ruled out by direct detection and/or BR($\mu\to e\gamma$).

If the scalar dark matter is dominated by the doublet component,
it couples also to the weak gauge bosons. LHC constraints then
come in addition from events with a single jet \cite{Aaboud:2017phn} and/or
vector boson \cite{Sirunyan:2017jix} and large transverse momentum
imbalance. They have only been interpreted for fermion dark
matter as a function of the mediator mass. If we take the spin to
be of minor importance and the mediator to be $Z$-like in coupling
and mass, the limits on the dark matter mass in Figs.\ 5 and 6 of
Ref.\ \cite{Aaboud:2017phn} and Fig.\ 10 of Ref.\ \cite{Sirunyan:2017jix}
lie at 50--100 GeV, so that the LHC would exclude only a few models
beyond the Higgs resonance region. These limits apply more directly
to the case of fermion dark matter, which we have not discussed.

\begin{figure}
\begin{center}
\begin{minipage}{0.45\textwidth}
\begin{picture}(300,100)(0,0)
   \put(10,0){\mbox{\resizebox{!}{0.5\textwidth}{\includegraphics{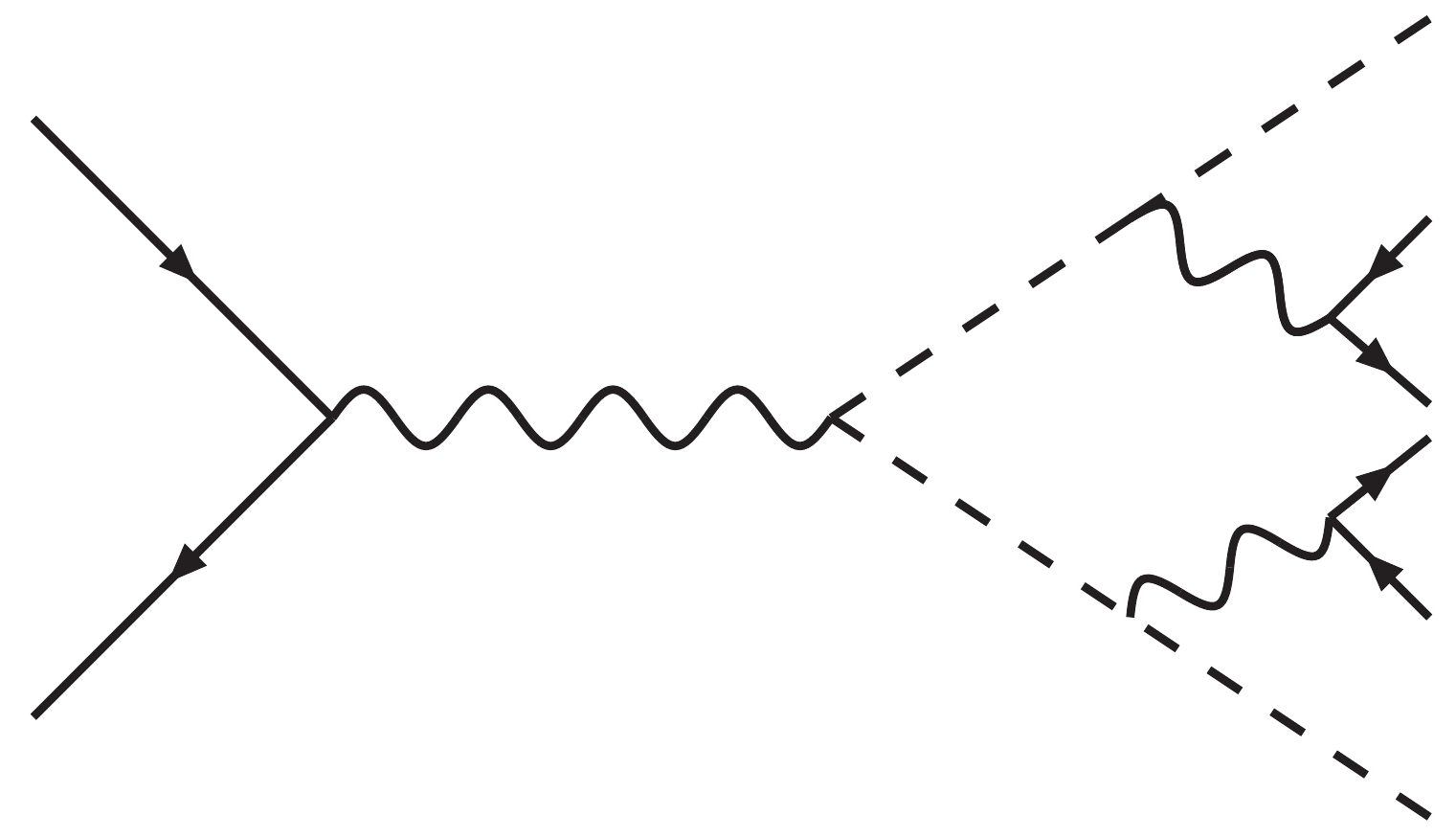}}}}
   \put(0, 2){$\bar{q}^{(\prime)}$}
   \put(0,90){$q$}
   \put(60,60){$Z(W^{\pm})$}
   \put(95, 80){$\phi^{\pm}$($X_{2,3}$)}  
   \put(95, 20){$\phi^{\mp}$($X_{2,3}$)}  
   \put(180,70){$l^\pm$}
   \put(180,55){$\nu$($l^\mp$)}
   \put(180,40){$l^\mp$}
   \put(180,20){$\nu$($l^\pm$)}
   \put(180,100){$X_1$ }
   \put(180, 0){$X_1$}
\end{picture}\end{minipage}\qquad
\begin{minipage}{0.45\textwidth}\begin{picture}(300,100)(0,0)
   \put(10,0){\mbox{\resizebox{!}{0.5\textwidth}{\includegraphics{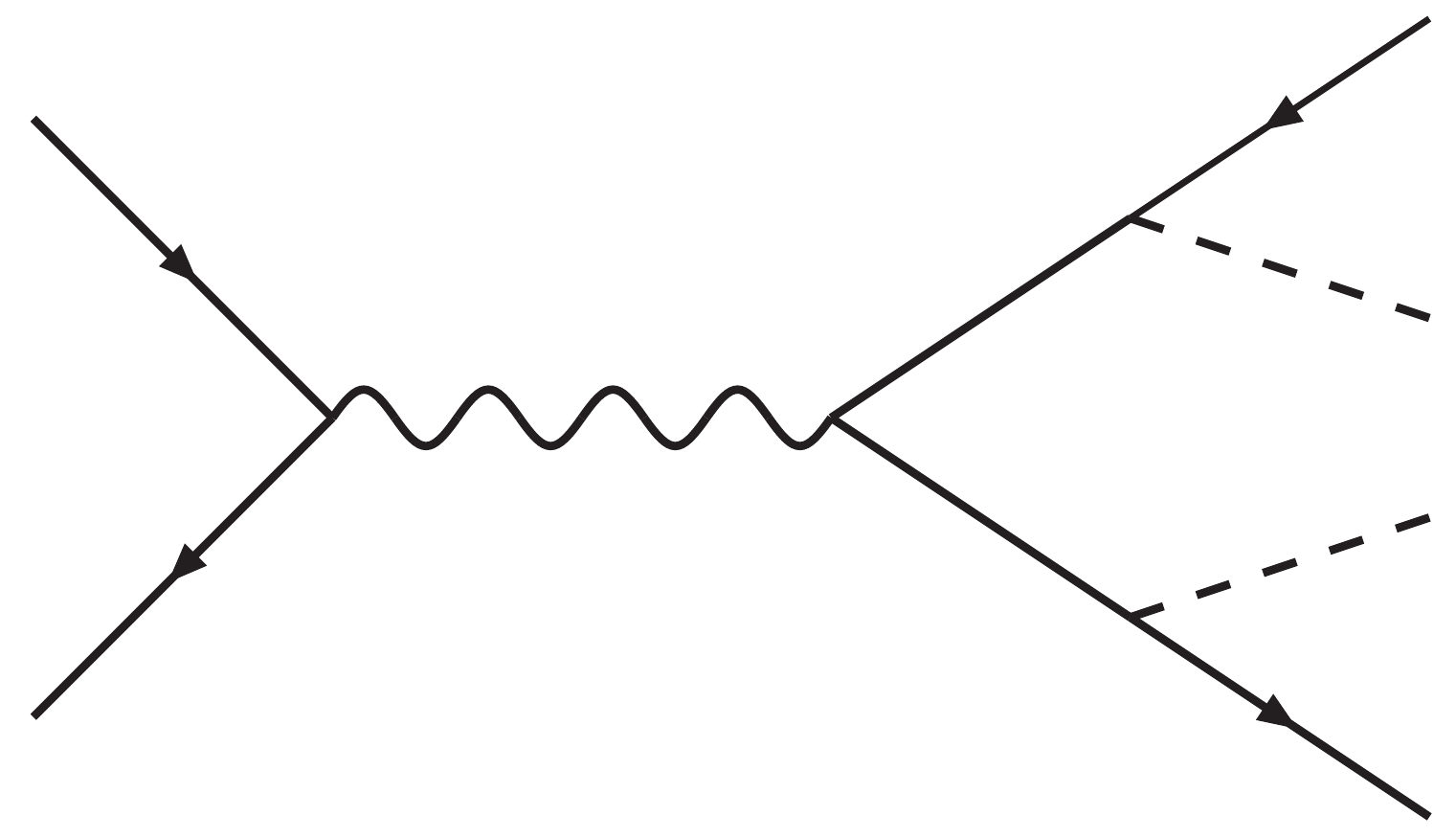}}}}
   \put(0, 2){$\bar{q}^{(\prime)}$}
   \put(0,90){$q$}
   \put(50,60){$Z(\gamma,W^{\pm}$)}
   \put(80,80){$\phi^\pm(X_{2,3})$}
   \put(80,20){$\phi^\mp(X_{2,3})$}
   \put(180,100){$l^\pm$($\nu$)}
   \put(180,2){$l^\mp$($\nu$)}
   \put(180,60){$X_1$ }
   \put(180,20){$X_1$}
\end{picture}\end{minipage}
\end{center}
\caption{Typical diagrams for the production of heavy scalars (left)
 and fermions (right) decaying into two- or multi-lepton final states
 and missing transverse energy, carried away by scalar dark matter
 $X_1$ and neutrinos.}
\label{fig:18}
\end{figure}
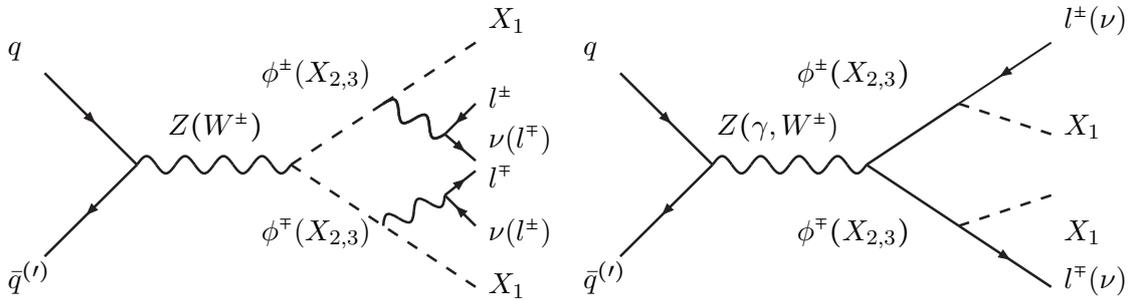

Alternatively, heavier charged and/or neutral scalars can be
produced at the LHC through weak gauge bosons. They decay subsequently
into dark matter and $W$- or $Z$-bosons, leading to two- or multi-lepton
signals with missing transverse energy (see Fig.\ \ref{fig:18} left)
\cite{Dolle:2009ft}.
These signals have been analysed at the LHC mostly in the context of
charged scalar leptons appearing in supersymmetric models. Scalar neutrinos
are so far unconstrained by the LHC. Although the charged sleptons do not translate
directly into our model, since they decay into fermionic neutralino dark matter
in supersymmetry, we can again assume spin to be of minor importance
and our charged scalars to decay into the lightest neutral scalars
and weak bosons with branching ratios of one, which leads to the strongest
constraints. Similar assumptions are also used in the slepton analyses
of ATLAS  \cite{ATLAS-CONF-2017-039} and CMS \cite{CMS-PAS-SUS-17-009},
which leads to lower slepton mass limits of 520 GeV in Fig.\ 6b of Ref.\
\cite{ATLAS-CONF-2017-039} and 440 GeV in Fig.\ 3 of Ref.\
\cite{CMS-PAS-SUS-17-009}. The corresponding mass limits for dark matter
then range from 50-280 GeV and 40-220 GeV,
respectively, which would in principle exclude some of the models beyond
the Higgs resonance region in Fig.\ \ref{fig:14}. The limits in our
model will, however, be considerable weaker since one has to take into account
in addition the leptonic branching fractions of the $W$- and $Z$-bosons
of 21\% and 7\% for electrons and muons, respectively.
If the charged scalars are heavier than 440-520 GeV or if they have other
(e.g.\ cascade) decays, the LHC limits of course no longer apply at all.

Doublet fermions, which coannihilate in our second random scan
with singlet dark matter, can be constrained in a similar way,
in particular from searches for higgsino-like charginos and
neutralinos (see Fig.\ \ref{fig:18} right) \cite{Restrepo:2015ura}. 
Their masses have been constrained by LEP and ATLAS in Fig.\ 10 of
Ref.\ \cite{Aaboud:2017leg} to be at least 95-145 GeV and by CMS
in Fig.\ 8 of Ref.\ \cite{Sirunyan:2018iwl} to be at least 100-170 GeV.
The fermionic neutralino dark matter mass then has to be 95-140 GeV
and 100-150 GeV, respectively. This would not affect many of our
viable models in Fig.\ \ref{fig:14} beyond the Higgs resonance region.
However, the limits in our model will here be stronger, since one does
not have into account the leptonic branching fractions of the $W$ and
$Z$-bosons. For singlet-doublet
fermion dark matter with scalar singlets, the charged fermions
have been shown to require masses of at least 510 GeV \cite{Restrepo:2015ura}. 

\section{Conclusion}
\label{sec:7}

In conclusion, we have analysed in this paper a combination of the
singlet-doublet scalar and singlet-doublet fermion dark matter
models, which had so far only been analysed separately. Their combination
allowed for the radiative generation of neutrino masses, but also
admitted lepton-flavour violating processes.

After discussing the
analytic structure and main parameter dependencies of the model and
the implications for the dark matter relic density, neutrino masses
and lepton flavour violation, we performed two random scans of the
parameter space, focusing on the case of scalar dark matter.

In the first scan over the full parameter space, we imposed contraints
from the Higgs mass, relic density and neutrino masses and mixings
using the Casas-Ibarra parameterisation. We found that the scalar dark
matter could be a mixture of singlets and doublets and in particular
that the scalar-fermion
couplings opened up large new regions of parameter space, mostly with
direct detection cross sections that will escape experimental verification
way beyond XENONnT. Many of these models were instead shown to be excluced
by lepton-flavour violation constraints. The remaining viable models
will soon be probed by XENONnT.

In the second scan, we focused on singlet scalar dark matter coannihilating
with mostly doublet fermions. In this case we also imposed constraints from
direct detection and lepton-flavour violation experiments. We found that at least one
of the scalar-fermion couplings and Yukawa couplings had to be large or not
too small, respectively, leading to two distinct regions in parameter space.
Many of these models will soon be tested by experiments on $\mu\to e\gamma$
or $\mu\to3e$, while those on $\tau\to e\gamma$ would require an increase
of at least two more orders of magnitude beyond current planning to allow
for the complete testing of at least one of the two viable regions in
parameter space.

LHC constraints from invisible Higgs decays were shown to add no further
constraints on the models with dark matter masses below the Higgs resonance,
while monojet, mono-boson and in particular dilepton searches have in
principle the potential to reach into the region beyond it up to 280 GeV.
These limits depend, however, crucially on the heavier and in particular
charged particle mass spectrum and decay modes. A full analysis of the
model would require detailed information on acceptances and efficiencies
of the LHC experiments and is beyond the scope of this work.

\begin{acknowledgments}
We thank K.\ Kovarik, S.\ May and M.\ Sunder for useful discussions.
This work has been supported by the DFG through the Research Training
Network 2149 ``Strong and weak interactions - from hadrons to dark matter''.
\end{acknowledgments}

\bibliographystyle{JHEP}
\bibliography{bib}

\end{document}